\documentclass{article}

\usepackage{arxiv}

\usepackage[utf8]{inputenc} % allow utf-8 input
\usepackage[T1]{fontenc}    % use 8-bit T1 fonts
\usepackage{hyperref}       % hyperlinks
\usepackage{url}            % simple URL typesetting
\usepackage{booktabs}       % professional-quality tables
\usepackage{amsfonts}       % blackboard math symbols
\usepackage{nicefrac}       % compact symbols for 1/2, etc.
\usepackage{microtype}      % microtypography
\usepackage{lipsum}
\usepackage{color}
\usepackage{algorithm}
\usepackage{algorithmic}
\usepackage{amsmath,amsfonts,amssymb}
\usepackage{subfigure}
\usepackage{lineno}
\usepackage{graphicx}
\usepackage{bm}
\graphicspath{{Figures/}}
\usepackage{subfigure}

\title{Super-Resolution Analysis via Machine Learning:
\\ A Survey for Fluid Flows}

\author{
Kai Fukami$^{[1,*]}$, Koji Fukagata$^{[2]}$, Kunihiko Taira$^{[1]}$\\
1. Department of Mechanical and Aerospace Engineering,
University of California, Los Angeles, CA 90095, USA\\
2. Department of Mechanical Engineering,
Keio University, Yokohama, 223-8522, Japan\\
Corresponding author: kfukami1@g.ucla.edu
}

\begin{document}
\maketitle

\begin{abstract}
This paper surveys machine-learning-based super-resolution reconstruction for vortical flows.
Super resolution aims to find the high-resolution flow fields from low-resolution data and is generally an approach used in image reconstruction.
In addition to surveying a variety of recent super-resolution applications, we provide case studies of super-resolution analysis for an example of two-dimensional decaying isotropic turbulence.
We demonstrate that physics-inspired model designs enable successful reconstruction of vortical flows from spatially limited measurements.
We also discuss the challenges and outlooks of machine-learning-based super-resolution analysis for fluid flow applications.
The insights gained from this study can be leveraged for super-resolution analysis of numerical and experimental flow data.
\end{abstract}

\setcounter{tocdepth}{5}
\tableofcontents

%%%%%%%%%%%%%%%%%%%%%%%%%%%%%%%%%%%%%%%%%%%%%%%%%%%%%%%%%%%%%%%%%%%%%%%%%%%
%%%%%%%%%%%%%%%%%%%%%%%%%%%%%%%% CHAPTER 1 %%%%%%%%%%%%%%%%%%%%%%%%%%%%%%%%
%%%%%%%%%%%%%%%%%%%%%%%%%%%%%%%%%%%%%%%%%%%%%%%%%%%%%%%%%%%%%%%%%%%%%%%%%%%
\section{Introduction}
\label{sec:intro}

Super resolution reconstructs a spatially high-resolution field data ${\bm q}_{\rm HR}$ from its low-resolution counterpart~${\bm q}_{\rm LR}$~\cite{irani1991improving,salvador2016example,bannore2009iterative}.
This problem set has been traditionally tackled in computer visions with various techniques including interpolation~\cite{keys1981cubic,vandewalle2006frequency,joshi2008psf,lucas1981iterative}, example-based internal learning~\cite{michaeli2013nonparametric,glasner2009super,zontak2013separating,shahar2011space}, high-frequency transfer~\cite{freedman2011image,yang2013fast,baker2002limits,park2003super}, neighbor embedding~\cite{roweis2000nonlinear,bevilacqua2012low,chang2004super,freeman2002example,freeman2000learning}, and sparse coding~\cite{lee2006efficient,yang2008image,lu2012geometry,yang2010image,zhang2012single}.
Although these implementations are effortless, it is generally challenging to reconstruct high-wavenumber contexts.
To address this difficulty, machine learning has been used for accurate super-resolution reconstruction of images~\cite{dong2014learning,dong2016accelerating,yang2019deep}. 
Machine learning can find a nonlinear relationship between input and output data even under ill-posed conditions.
This approach can be applied to a pair of low- and high-resolution images, providing a finer level of images from extremely coarse images~\cite{dong2015image}.

Machine-learning-based techniques in general~\cite{BNK2020,BHT2020,BEF2019} have been considered for a range of applications in fluid mechanics including turbulence modeling~\cite{DIX2019,MSJC2019,LKT2016,novati2021automating,bae2022scientific}, reduced-order modeling~\cite{LY2019,callaham2022empirical,san2018neural,fukami2020sparse,SGASV2019}, data reconstruction~\cite{RPCA2020,manohar2018data,FFT2020,KKL2023}, and flow control~\cite{rabault2019artificial,bieker2020deep,zhou2020artificial,paris2021robust,park2020machine,ghraieb2021single}.
Super-resolution reconstruction with machine learning is no exception.
The lower barrier to access open source codes in image science and implement models also enables fluid mechanicians to apply methods for fluid flow data by replacing RGB components (red, green, and blue) with velocity components~$\{u,v,w\}$.

While super resolution can be regarded as an image-based data recovery technique, it is also a general framework for a broad range of applications in fluid mechanics.
For instance, a low-resolution fluid flow image can be interpreted as a set of sparse sensor measurements. 
In this aspect, the inverse problem of global field reconstruction from local measurements is an extension of super-resolution analysis~\cite{fukami2021global,guemes2022super,sun2020physics}.
If we consider low-resolution fluid flow data as noisy experimental measurements, super-resolution analysis can also be extended to denoising problem~\cite{gao2021super,fathi2020super,vlasenko2009superresolution}.
Furthermore, large-eddy simulation (LES) can incorporate super-resolution reconstruction to reveal finer structures inside a low-resolution grid cell~\cite{pradhan2021variational,bode2021using}.

This paper surveys the current status and the challenges of machine-learning-based super-resolution analysis for vortical flows.
We first cover several machine-learning models and their applications to super resolution of fluid flows.
We then offer case studies using a supervised learning-based super resolution for an example of two-dimensional decaying isotropic turbulence.
We consider embedding physics into the model design to successfully reconstruct a high-resolution vortical flow from low-resolution data.
We further discuss the challenges and outlooks of machine-learning-based super resolution in fluid flow applications.
The present paper is organized as follows.
We introduce machine-learning approaches of super-resolution reconstruction for vortical flows in section~\ref{sec:methods}.
Applications of these machine-learning techniques are discussed in section~\ref{sec:appl}.
We perform case studies in section~\ref{sec:results}.
Extensions of super-resolution analysis for fluid dynamics are discussed in section~\ref{sec:outlook}. 
Concluding remarks with outlooks are provided in section~\ref{sec:conc}.

\section{Approaches}
\label{sec:methods}

A variety of machine-learning models have been proposed for the super-resolution reconstruction of vortical flows, as summarized in table~\ref{tab1}.
% While traditional interpolations are unable to recover finer signal information than existing scales, 
Machine-learning-based approaches can find a nonlinear relationship between the low-resolution input and the corresponding high-resolution output from a large collection of data through training.
In super-resolution analysis, the dimension of the input (low-resolution data) ${\bm q}_{\rm LR} \in \mathbb{R}^{m}$ is smaller than that of the high-resolution output ${\bm q}_{\rm HR} \in \mathbb{R}^{n}$ with $m\ll n$,
\begin{align}
    {\bm q}_{\rm HR} = F({\bm q}_{\rm LR}),
\end{align}
where $F$ is the super-resolution model.
Depending on the flow of interest and the size of data, the machine-learning model should be carefully chosen. 
In section~\ref{sec:models}, we introduce three types of machine-learning models that are widely used. 
We also discuss the use of physics-based loss functions in section~\ref{sec:PILF}.

\begin{table}[H]
\caption{Representative studies on machine-learning-based super-resolution reconstruction methods for fluid flows.}
\label{tab1}
\centering
\begin{tabular}{c|c|c|c|c}
\hline\hline
Authors (Year) & Flow examples & Approaches & Algorithms & Notes  \\ \hline

\begin{tabular}[c]{@{}c@{}}Xie, Franz, Chu,\\  Thuerey (2018)~\cite{xie2018tempogan}\end{tabular}    
& Smoke dynamics 
& Supervised 
& \begin{tabular}[c]{@{}c@{}}MLP/CNN\\ (tempoGAN)\end{tabular} 
& \begin{tabular}[c]{@{}c@{}}Incorporates dynamics, \\ expensive\end{tabular}  
\\ \hline

\begin{tabular}[c]{@{}c@{}}Fukami, Fukagata,\\ Taira (2019)~\cite{FFT2019a}\end{tabular}    
& \begin{tabular}[c]{@{}c@{}}Laminar cylinder wake,\\ isotropic turbulence, \\ turbulent channel flow\end{tabular} 
& Supervised 
& \begin{tabular}[c]{@{}c@{}}CNN\\ (DSC/MS)\end{tabular} 
& \begin{tabular}[c]{@{}c@{}}Robust for \\ multi-scale physics, \\ stability of training\end{tabular} 
\\ \hline

\begin{tabular}[c]{@{}c@{}}Erichson, Mathelin, \\  Yao, Brunton, Mahoney,\\ Kutz (2020)~\cite{erichson2020shallow} \end{tabular} 
& \begin{tabular}[c]{@{}c@{}}Laminar cylinder wake,\\ isotropic turbulence, \\ sea surface temperature\end{tabular} 
& Supervised 
& MLP 
& \begin{tabular}[c]{@{}c@{}}Accuracy, simplicity, \\ expensive\end{tabular} 
\\ \hline

\begin{tabular}[c]{@{}c@{}}Bode, Gauding, \\  Kleinheinz, Pitsch \\(2019)~\cite{bode2019deep} \end{tabular}    
& Isotropic turbulence
& Unsupervised 
& \begin{tabular}[c]{@{}c@{}}CNN\\ (ESRGAN)\end{tabular} 
& \begin{tabular}[c]{@{}c@{}}Physics loss function, \\stability of training\end{tabular} 
\\ \hline

\begin{tabular}[c]{@{}c@{}}Obiols-Sales, \\ Vishnu, Malaya, \\ Chandramowlishwaran\\ (2021)~\cite{obiols2021surfnet}\end{tabular}  
& \begin{tabular}[c]{@{}c@{}}NACA airfoil wake, \\ cylinder wake\end{tabular} 
& Supervised 
& \begin{tabular}[c]{@{}c@{}}CNN\\ (SURFNet)\end{tabular} 
& \begin{tabular}[c]{@{}c@{}}Uncertainties\\ over transfer,\\ inexpensive
\end{tabular}
\\ \hline

\begin{tabular}[c]{@{}c@{}}Liu, Tang, Huang, \\ Lu (2020)~\cite{liu2020deep}\end{tabular}  
& \begin{tabular}[c]{@{}c@{}}Isotropic turbulence, \\ turbulent channel flow\end{tabular} 
& Supervised 
& \begin{tabular}[c]{@{}c@{}}CNN \\ (MTPC)\end{tabular} 
& \begin{tabular}[c]{@{}c@{}}Incorporates dynamics, \\ high-wavenumber\\reconstruction\end{tabular}
\\ \hline

\begin{tabular}[c]{@{}c@{}}Kim, Kim, Won, \\ Lee (2021)~\cite{kim2021unsupervised}\end{tabular}  
& \begin{tabular}[c]{@{}c@{}}Isotropic turbulence, \\ turbulent channel flow\end{tabular} 
& Unsupervised 
& \begin{tabular}[c]{@{}c@{}}CNN\\ (Cycle GAN)\end{tabular} 
& \begin{tabular}[c]{@{}c@{}}No required\\ paired data,\\ stability of training\end{tabular} 
\\ \hline

\begin{tabular}[c]{@{}c@{}}Gao, Sun, Wang \\ (2021)~\cite{gao2021super}\end{tabular}  
& \begin{tabular}[c]{@{}c@{}} Laminar flow with \\ spatially varying BCs, \\ cardiovascular flow\end{tabular} 
& Semi-supervised 
& CNN 
& \begin{tabular}[c]{@{}c@{}}Less data, \\ physics loss function, \\tuning of\\ loss coefficients \end{tabular}
\\ \hline

\begin{tabular}[c]{@{}c@{}}Zhou, McClure, \\Chen, Xiao (2022)~\cite{zhou2022neural}\end{tabular}  
& Porous flow 
& Supervised 
& \begin{tabular}[c]{@{}c@{}}CNN\\ (U-Net)\end{tabular} 
& \begin{tabular}[c]{@{}c@{}}Robust against noise, \\ need to compress data \end{tabular}
\\ \hline

\begin{tabular}[c]{@{}c@{}}G{\"u}emes, Discetti, \\ Ianiro, Sirmacek, \\ Azizpour, Vinuesa\\ (2021)~\cite{guemes2021coarse}\end{tabular}
& Turbulent channel flow
& Unsupervised 
& \begin{tabular}[c]{@{}c@{}}CNN\\ (SRGAN)\end{tabular}
& \begin{tabular}[c]{@{}c@{}}Small-scale \\ reconstruction,\\ training data size \end{tabular}
\\ \hline

\begin{tabular}[c]{@{}c@{}}Pradhan, Duraisamy \\ (2021)~\cite{pradhan2021variational}\end{tabular}  
& Turbulent channel flow 
& Supervised 
& \begin{tabular}[c]{@{}c@{}}MLP\\ (VSRNN)\end{tabular} 
& \begin{tabular}[c]{@{}c@{}}Generalized for \\ unseen initial condition, \\ expensive\end{tabular}
\\ \hline

\begin{tabular}[c]{@{}c@{}}Fukami, Maulik,\\Ramachandra, \\Fukagata, Taira\\ (2021)~\cite{fukami2021global}\end{tabular}
& \begin{tabular}[c]{@{}c@{}}Laminar cylinder wake,\\ sea surface temperature,\\ turbulent channel flow \end{tabular}
& Supervised 
& \begin{tabular}[c]{@{}c@{}}CNN\\ (Voronoi model)\end{tabular} 
& \begin{tabular}[c]{@{}c@{}}Generalized for \\ moving sensors, \\ training data size \end{tabular}
\\ \hline

\begin{tabular}[c]{@{}c@{}}Yousif, Yu, Lim \\ (2021)~\cite{yousif2021high}\end{tabular}
& Turbulent channel flow
& Unsupervised 
& \begin{tabular}[c]{@{}c@{}}CNN\\ (MS-ESRGAN)\end{tabular} 
& \begin{tabular}[c]{@{}c@{}}Robust for\\ multi-scale physics, \\ expensive\end{tabular} 
\\ \hline

\begin{tabular}[c]{@{}c@{}}Bode, Gauding,\\ Lian, Denker, \\ Davidovic, Kleinheinz, \\Jitsev, Pitsch \\ (2021)~\cite{bode2021using}\end{tabular}
& Isotropic turbulence 
& Unsupervised 
& \begin{tabular}[c]{@{}c@{}}CNN\\ (PIESRGAN)\end{tabular} 
& \begin{tabular}[c]{@{}c@{}} Physics loss function, \\tuning of \\ loss coefficients\end{tabular}
\\ \hline

\begin{tabular}[c]{@{}c@{}}Nair, Goza (2020)~\cite{NG2020}\end{tabular}  
& Laminar flat plate wake 
& Supervised 
& MLP 
& \begin{tabular}[c]{@{}c@{}} Inexpensive, \\ requires POD\end{tabular}\\
\hline\hline
\end{tabular}
\end{table}

\subsection{Machine-learning models}
\label{sec:models}

\subsubsection{Fully-connected network (multi-layer perceptron)}

%% FIGURE 1 %%%%%%%%%%%%%%%%%%%%%%%%%%%%%%%%%%%%%%%%%%%
\begin{figure*}
    \centering
		%\hspace{-30mm}
		\includegraphics[width=0.75\textwidth]{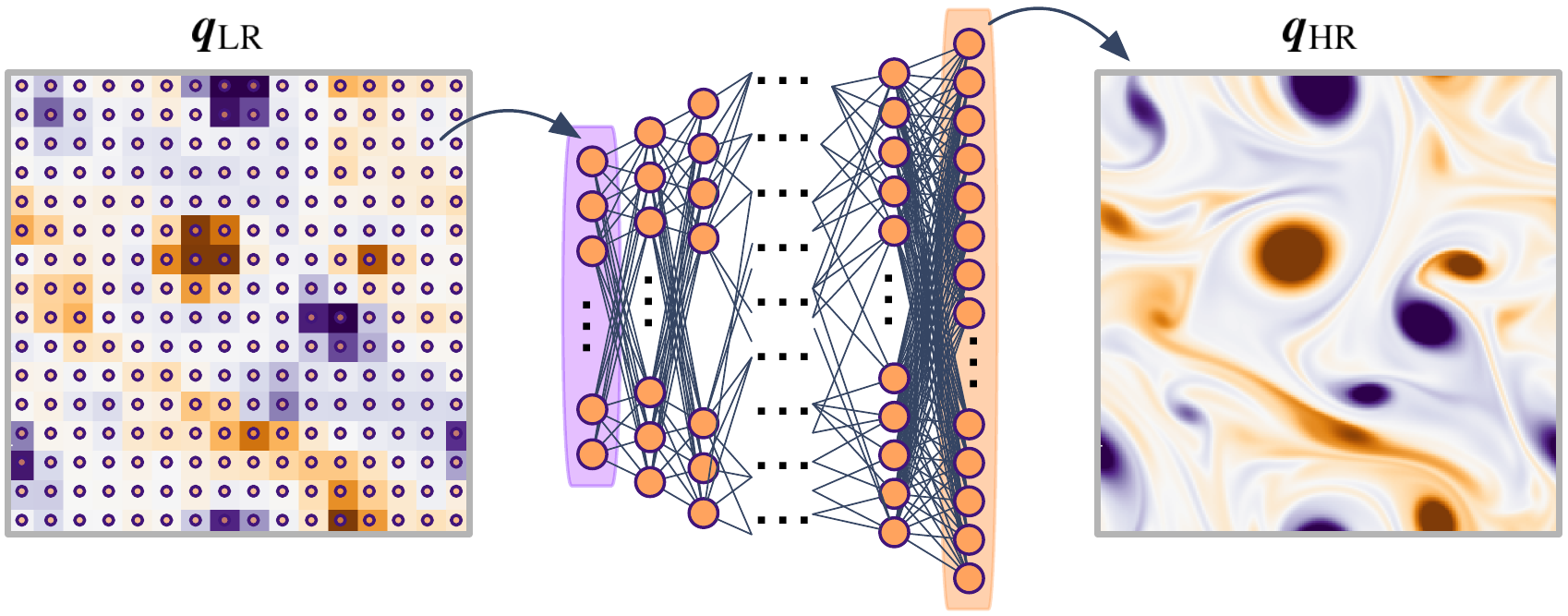}
		\caption{Fully-connected model-based super resolution.}
		\label{fig_MLP}
\end{figure*}
%%%%%%%%%%%%%%%%%%%%%%%%%%%%%%%%%%%%%%%%%%%%%%%%%%%%%%

The fully-connected network, also called the multi-layer perceptron~\cite{RHW1986}, is the most basic neural network model.
Nodes between layers are fully connected with each other, as illustrated in figure~\ref{fig_MLP}.
The minimum unit of a fully-connected network is called perceptron.
For each perceptron, the linear combination of the inputs from layer $(l-1)$, $c_j^{(l-1)}$, is connected with weights~${\bm w}$ yielding the output at layer $(l)$, $c_i^{(l)}$,
\begin{equation}
    {c}^{(l)}_{i}=\varphi(\sum_{j}{w}_{ij}^{(l)}{c}^{(l-1)}_{j} + b_i^{(l)}),    
\end{equation}
where $\varphi$ is the activation function and $b$ is the bias added at each layer.
We can choose a nonlinear function for $\varphi$, enabling the network to capture the nonlinear relationship between the input and the output.

A fully-connected model can be used for supervised machine learning-based super resolution.
A training process for supervised machine-learning models is cast as an optimization problem to determine the weights~${\bm w}$ inside the model $F$. 
The weights ${\bm w}$ are optimized by minimizing the loss function~$\cal{E}$ through backpropagation~\cite{Kingma2014}.
This optimization procedure is described as
\begin{equation}
    {\bm w}={\rm argmin}_{\bm w}~{\cal E}({\bm w}).
    \label{eq2}
\end{equation}
Since super-resolution reconstruction aims to obtain a high-resolution image ${\bm q}_{\rm HR}$ from the corresponding low-resolution data ${\bm q}_{\rm LR}$, the loss function (error) can be formulated as 
\begin{align}
    {\cal E} &= ||{\bm q}_{\rm HR}-F({\bm q}_{\rm LR})||_P,~\label{eq_super}
\end{align}
where $P$ indicates the norm.
While the $L_2$ norm is widely used, we can instead consider other norms such as $L_1$ norm and logarithmic norm depending on the data characteristics.
The $L_1$ norm can be used for model construction that is not as sensitive for outliers in the data. 
The logarithmic norm is suitable for cases where underestimation should be avoided.

As mentioned above, the difference of data dimension between the input and the output in the super-resolution analysis is substantial.
Hence, models generally comprise the decoder-type structure~\cite{erichson2020shallow,williams2022data}, meaning that the number of nodes gradually increases towards the output layer.
This is especially the case for high-dimensional inverse problems such as super-resolution reconstruction of fluid flows.
This leads to the number of nodes and their connections to drastically increase, leading to the prohibitively expensive computational cost and the failure of non-convex optimization known as the curse of dimensionality~\cite{domingos2012few}.
Users should be mindful of computational time and memory requirements for fully-connected models.

\subsubsection{Convolutional neural network}

To address the issue of the computational burden associated with the fully-connected models, convolutional neural networks (CNNs)~\cite{LBBH1998} have been widely utilized in super-resolution analysis of fluid flows.
CNNs incorporate a function called filter sharing, enabling the processing of large vortical flow data without encountering the curse of dimensionality~\cite{morimoto2021convolutional}.

\begin{figure*}
    \centering
		%\hspace{-30mm}
		\includegraphics[width=0.7\textwidth]{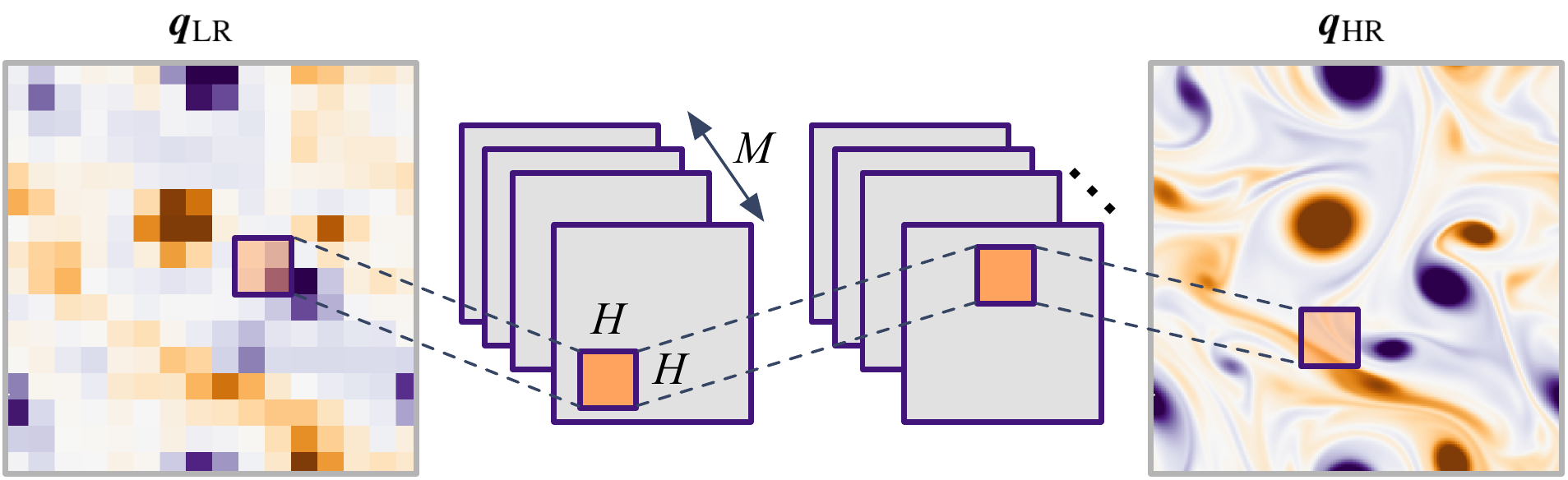}
		\caption{Convolutional neural network-based super resolution.}
		\label{fig_CNN}
\end{figure*}

A CNN is generally comprised of the convolutional layer, pooling layer, and upsampling layer.
The convolutional layer depicted in figure~\ref{fig_CNN} captures the nonlinear relationship between input and output data by extracting spatial features of supplied data through filtering operations.
This operation is expressed as
\begin{equation}
    q^{(l)}_{ijn}=\varphi\left(\sum_{m=1}^M\sum_{p=0}^{H-1}\sum_{q=0}^{H-1}h^{(l)}_{pqmn}q^{(l-1)}_{i+p-G,j+q-G,m}+b_n^{(l)}\right),
    \label{eq:CNN}
\end{equation}
where $G=\lfloor H/2\rfloor$, $H$ is the width and height of the filter, $M$ is the number of input channel, $n$ is the number of output channel, $b$ is the bias, and $\varphi$ is the activation function.
As in the fully-connected models, a nonlinear function can be chosen for $\varphi$ to account for nonlinearlities in the machine-learning model.

%% FIGURE 1 %%%%%%%%%%%%%%%%%%%%%%%%%%%%%%%%%%%%%%%%%%%
%%%%%%%%%%%%%%%%%%%%%%%%%%%%%%%%%%%%%%%%%%%%%%%%%%%%%%

In addition to the convolutional layer, a pooling layer also plays an important role in CNN-based analysis.
The pooling layer downscales the data, reducing data dimension.
For regression tasks, it is useful for reducing spatial sensitivity, producing a robust CNN model against noisy inputs~\cite{NF2022}.
It is also possible to expand the data dimension through the upsampling layer.
Upsampling copies the value onto an arbitrary region to expand the dimension.
This function is especially useful to align the data dimension inside the network.

For super resolution in which the dimension of the output~$\mathbb{R}^{n}$ is larger than that of the input~$\mathbb{R}^{m}$, there are several ways to treat the difference of the dimensions between the input and the output.
For example, the upsampling can be used inside a network to expand the dimension~\cite{wurster2022deep}.
One can also implement a resize or interpolation function for the input data to align the size with that of the output~\cite{dong2015image,FFT2019a,romano2016raisr}.
This can avoid the use of pooling or upsampling operations, reducing the complexity of the model.

\subsubsection{Generative adversarial network}
\label{sec:GAN}

%% FIGURE 1 %%%%%%%%%%%%%%%%%%%%%%%%%%%%%%%%%%%%%%%%%%%
\begin{figure*}[b]
    \centering
		%\hspace{-30mm}
		\includegraphics[width=0.8\textwidth]{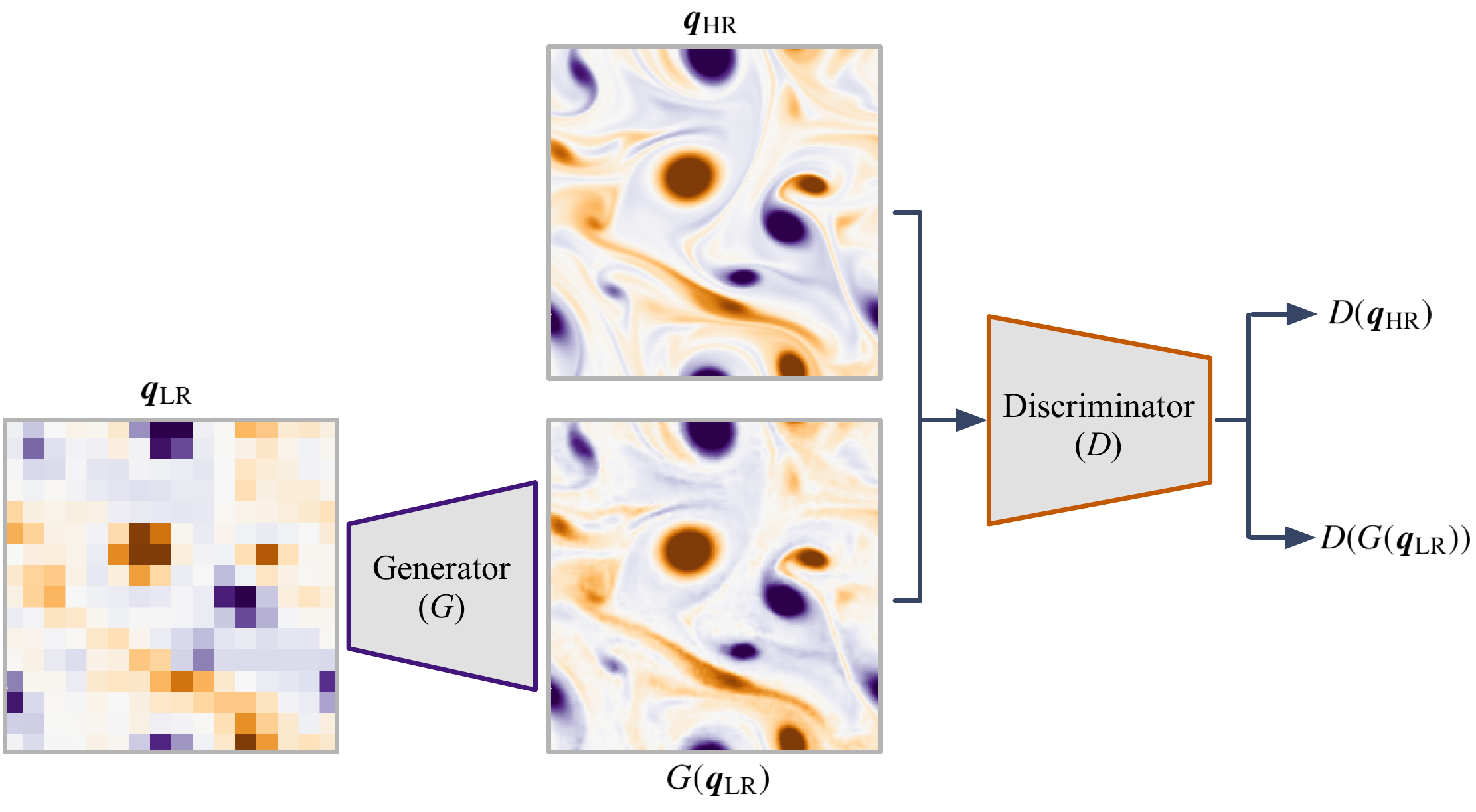}
		\caption{Generative adversarial network-based super resolution.}
		\label{fig_GAN}
\end{figure*}
%%%%%%%%%%%%%%%%%%%%%%%%%%%%%%%%%%%%%%%%%%%%%%%%%%%%%%

In addition to supervised fully-connected networks and convolutional networks, unsupervised learning with generative adversarial network (GAN)~\cite{goodfellow2020generative} has also been proposed for super-resolution analysis of fluid flows~\cite{xie2018tempogan,bode2019deep,kim2021unsupervised,guemes2021coarse,MTK2023}.
GAN is attractive for cases in which it is difficult to prepare paired input and output data.
For example, the application of super resolution with LES can correspond to this scenario.
A model trained with a pair of high-fidelity DNS and subsampled low-resolution data may not directly support super-resolution reconstruction for LES data.
% in the absence of corresponding DNS data.
Super resolution of PIV measurements with limited spatio-temporal resolution (without corresponding high-resolution solution images) also needs to be carefully considered.

GAN is composed of two networks, namely, a generator ($G$) and a discriminator ($D$).
A generator produces a fake image which is similar to the solution from random noise $\bm n$.
In contrast, a discriminator judges the generated (fake) image as whether it is likely to be a realistic image by returning a probability between 0 (fake) and 1 (real).
A generator usually possesses a decoder-type structure to expand the data dimension from noise to images, while a discriminator is composed of an encoder-type network towards reducing the data size from images to the probability.
Throughout the training process, the weights inside the generator are being updated to deceive the discriminator toward the direction of minimizing the probability by generating images increasingly similar to the real data.
Fake images produced by the generator eventually become high-quality images that cannot be distinguished from the real image.

These processes can be mathematically expressed with regard to the cost function $V(D,G)$,
\begin{align}
    \underset{G}{\rm min} \underset{D}{\rm max}~V(D,G) = \mathbb{E}_{{\bm d} \sim p_{\rm data}({\bm d})}[{\rm log}D({\bm d})]
    + \mathbb{E}_{{\bm n} \sim p_{\bm n}({\bm n})}[{\rm log}(1-D(G(\bm n)))],~\label{eq_GAN}
\end{align}
where $\bm d$ is a real data set and $p_{\rm data}$ is the probability distribution of the real data.
The parameters in the generator $G$ are trained towards the direction in which $D(G({\bm n}))$ becomes 1.
On the other hand, the weights in the discriminator $D$ are updated so that $D({\bm d})$ returns a value close to 1. 
Since the discriminator becomes wiser through training, $D(G({\bm n}))$ provides a value close to 0.
Summarizing, the parameter inside the generator $G$ is optimized by minimizing the loss function while that for the discriminator $D$ is adjusted by maximizing the loss function, referred to as competitive learning~\cite{rumelhart1985feature}.
Once the training ends, the trained generator can produce an output with indistinguishable quality compared to the real data.
For super-resolution problems, we can use low-resolution data as the input for the generator $G$ instead of random noise $\bm n$, as illustrated in figure~\ref{fig_GAN}.
A generator in super-resolution reconstruction provides a statistically plausible high-resolution output by learning the relationship between the input low-resolution data set and the high-resolution data set, which need not be paired.

\subsection{Choice of loss function}
\label{sec:PILF}

Here, let us discuss the choice of loss (cost) function for machine-learning-based super-resolution analysis.
In standard formulation, we can have the cost function defined by equations~\ref{eq_super} and~\ref{eq_GAN}. 
However, super-resolved flow fields with direct applications of machine-learning models do not satisfy physical conditions, such as the conservation laws.
To address such an issue, loss functions that embed physics laws can be utilized~\cite{lagaris1998artificial,raissi2019physics}.
Together with the original data-based cost ${\cal E}_d$ from equation~\ref{eq_super} or~\ref{eq_GAN}, the loss function ${\cal E}$ incorporating a physics-inspired loss function ${\cal E}_p$ for super-resolution analysis can take the form of
\begin{align}
    {\cal E} &= {\cal E}_d + \beta{\cal E}_p,
\end{align}
where $\beta$ provides a scale between ${\cal E}_d$ and ${\cal E}_p$.

There are several approaches to introduce the physics-based loss term for fluid flows.
For instance, we can directly substitute a reconstructed high-resolution field ${\bm q}_{\rm HR}$ into the governing equation~\cite{raissi2019physics} if we have all data for the state variables to have,
\begin{align}
    {\cal E}_p = ||{\cal N}({\bm x}, {\bm q}_{\rm HR}({\bm x},t))||_P,
\end{align}
where ${\cal N}$ is an operator from governing equations.
Minimizing a loss function incorporating only certain terms of the Navier--Stokes equation~\cite{LY2019} can also be considered,
\begin{align}
    &{\cal E}^j_p = ||{\cal N}_j({\bm q}_{\rm Ref})-{\cal N}_j({\bm q}_{\rm HR})||_P,~~~{\cal N} = \sum_j {\cal N}_j,
\end{align}
where ${\cal N}_j$ is a term in the governing equation and ${\bm q}_{\rm Ref}$ is a reference data.
It is known that these physics-based loss functions help in reconstructing flows with a small amount of data~\cite{gao2021super}.
What these terms in the loss function do is to better constrain the solution space~\cite{karniadakis2021physics,cai2022physics}.
This is a similar concept to semisupervised learning which combines a small amount of labeled data with a large amount of unlabeled data~\cite{zhu2009introduction}.
In the present paper, we demonstrate the effectiveness of training with small data set for super-resolution reconstruction of turbulent vortices in section~\ref{sec:results}.
We should however note that the so-called physics-inspired analysis can suffer from large numerical error if ${\bm q}_{\rm HR}$ contains error or noise.
This approach should be used with caution as it assumes that ${\bm q}_{\rm HR}$ can be used to evaluate certain terms.

\section{Applications}
\label{sec:appl}

In this section, we survey recent super-resolution applications for fluid flows through supervised (section~\ref{sec:slbt}) and semisupervised-/unsupervised learning (section~\ref{sec:sul}).

\subsection{Supervised learning}
\label{sec:slbt}

In machine-learning-based super-resolution reconstruction of fluid flows, supervised techniques are often used.
Supervised learning requires a pair of input and output flow field data as training data.
For super-resolution analysis, a high-resolution reference flow field and the corresponding low-resolution data need to be available for training models.
To avoid the curse of dimensionality, CNN models are often used for image-based super resolution of fluid flows rather than fully-connected models.

Fukami et al.~\cite{FFT2019a,FFT2019tsfp,FFT2021b} proposed a CNN-based super-resolution reconstruction for fluid flows in a supervised manner.
The CNN-based model was applied to examples of a two-dimensional cylinder wake, two-dimensional isotropic turbulence, and three-dimensional turbulent channel flow.
To capture multi-scale physics in turbulent vortical flows, they also proposed the hybrid downsampled skip-connection/multi-scale (DSC/MS) model based on the CNN.
The model is composed of the up-/downsampling operations, the skip connection~\cite{he2016deep}, and CNNs with various sizes of filters.
While up-/downsampling operations support robustness against rotation and translation of vortical structures, the skip connection provides stability of the learning process~\cite{he2016deep}.
Moreover, the multi-scale CNN aims to capture a variety of length scales in turbulent flows.
Especially for the examples of turbulence, it was shown that the DSC/MS model is effective in accurately preserving the energy spectrum.

\begin{figure*}[b]
    \centering
		%\hspace{-30mm}
		\includegraphics[width=0.85\textwidth]{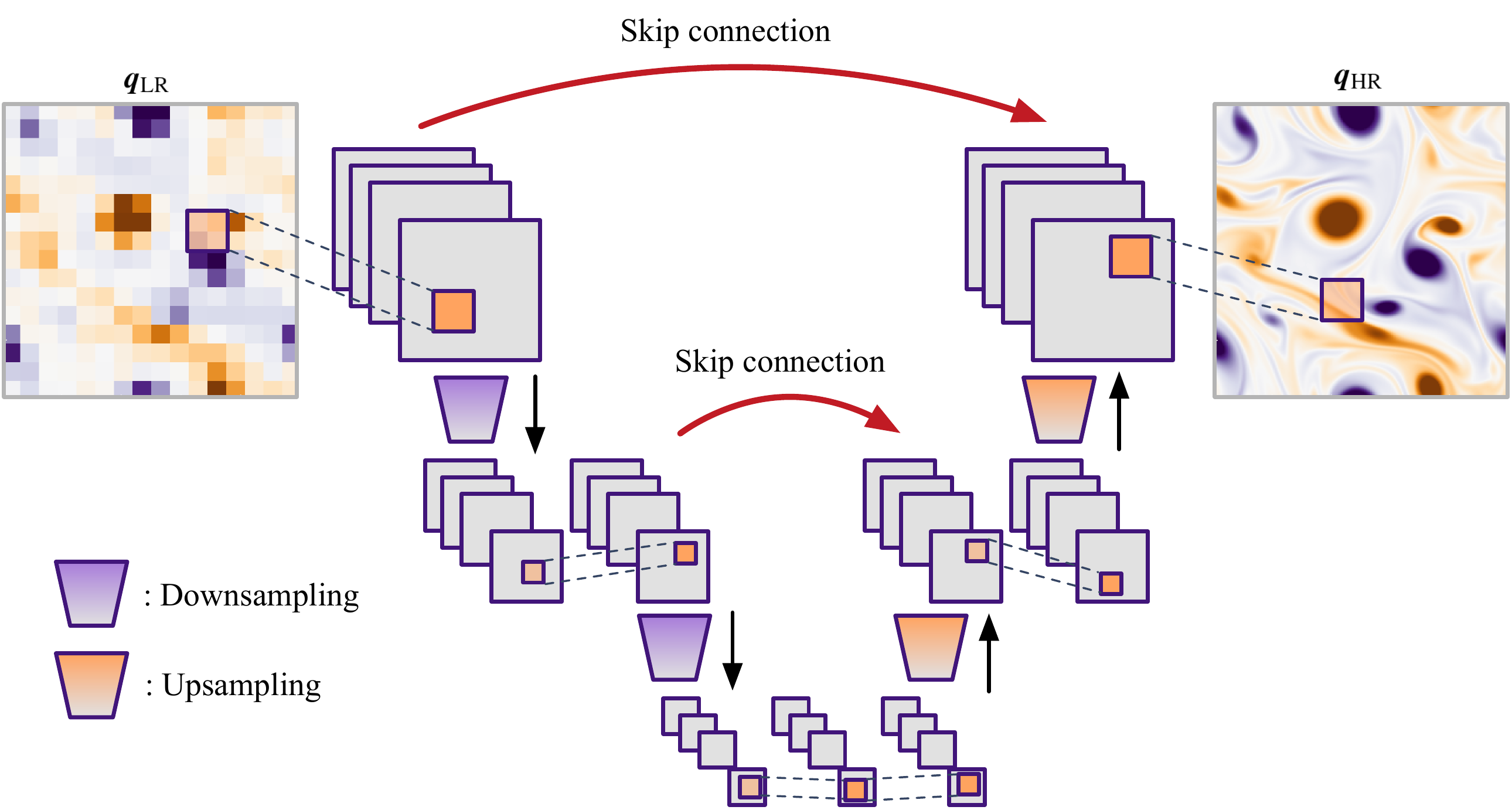}
		\caption{U-Net-based model for super-resolution reconstruction of vortical flows.}
		\label{fig_Unet}
\end{figure*}

Following this study, supervised CNN-based super-resolution analysis has been actively studied for a range of flows.
Obiols-Sales et al.~\cite{obiols2021surfnet} proposed a CNN-based super-resolution model called SURFNet and tested its performance for wakes around various NACA-type airfoils, ellipses, and cylinders.
SURFNet includes a transfer learning-based augmentation~\cite{pan2009survey}.
The model is first trained using only low-resolution flow data, and then the pre-trained weights are transferred in training with high-resolution data sets.
Transfer learning over multiple levels of spatial-resolution flow field can improve the accuracy of super-resolution reconstruction~\cite{guastoni2021convolutional}, which is also related to multi-fidelity learning~\cite{LPBK2020}.
% Wang et al.~\cite{wang2021novel} extended the DSC/MS model~\cite{FFT2019a} to reduce the training costs.
% Their network called enhanced DSC/MS model was tested for velocity fields around a cylinder and a hydrofoil.
% They also reported that the model can be universally applied to both wake fields with less computational costs compared to the original model.
U-Net-based model (illustrated in figure~\ref{fig_Unet}) can also reduce the training cost for super-resolution reconstruction of turbulent flows since the size of fluid flow data is reduced through an autoencoder-type model structures~\cite{pant2020deep}.
% A similar U-Net model was also applied to super resolution of porous flow by Zhou et al.~\cite{zhou2022neural}.
% Their U-Net model takes geometry information as the input and incorporates a coarse velocity field as supplemental information to improve the accuracy of estimation for high-resolution velocity fields.

%% FIGURE 1 %%%%%%%%%%%%%%%%%%%%%%%%%%%%%%%%%%%%%%%%%%%
%%%%%%%%%%%%%%%%%%%%%%%%%%%%%%%%%%%%%%%%%%%%%%%%%%%%%%

Incorporating physical insights and domain knowledge into model construction further supports or enhances supervised-learning-based super-resolution reconstruction in vortical flows.
For instance, accounting for spatial length scales of the flow structures in the models improves reconstruction~\cite{FFT2019a}.
Kong et al.~\cite{kong2020deep} developed a multiple path super-resolution CNN with several connections inside the model to capture variations of spatial temperature distribution in a supersonic combustor. 
They reported that the proposed multiple-path CNN provides enhanced reconstruction of temperature fields compared to a regular CNN.
Incorporating the time history of flow fields is also useful for super-resolving vortical flows in a supervised manner.
Liu et al.~\cite{liu2020deep} compared two types of supervised CNN-based models for super-resolution analysis: namely the static CNN (SCNN) and the multiple temporal paths CNN (MTPC). 
While the SCNN model uses instantaneous flow snapshots as the input, the MTPC model considers a time series of velocity fields as the input to read spatial and temporal information simultaneously.
With examples of forced isotropic turbulence and turbulent channel flow, they found that the MTPC model can improve the reconstruction of turbulence statistics such as kinetic energy spectra and the second and third invariants of the velocity gradient tensor.

Once supervised models are trained, machine-learning models can be used for data compression since we only need to save only the input data to recover high-resolution flow fields.
Matsuo et al.~\cite{matsuo2021supervised} proposed an adaptive super-resolution analysis.
They focused on how a low-resolution field is prepared in training a supervised learning-based model. 
While max- and averaging pooling operations are generally used for preparing low-resolution data sets, they considered the spatial standard deviation in arbitrary subdomains in a flow field to determine the local degree of downsampling.
This can account for the importance of flow structures in generating low-resolution data sets.
They reported that supervised CNN models can reconstruct a high-resolution field of three-dimensional square cylinder wake from adaptive low-resolution data, achieving approximately 0.05\% data compression against the original data.

Compressing fluid flow data in the time direction can also be considered.
Fukami et al.~\cite{FFT2021b} used the DSC/MS model to reconstruct high-resolution turbulent flows from coarse flow data in space and time inspired by a concept of super-resolution analysis and inbetweening~\cite{LRT2019}.
In their formulation, two spatial coarse flow fields at $t=n\Delta t$ and $t=(n+k)\Delta t$ are taken as the input of the first machine-learning model.
Once the spatial-reconstruction model provides two super-resolved high-resolution flow fields, these outputs are then fed into the second model to perform inbetweening that provides high-resolution snapshots between the beginning and the end frames.
By combining these two models, spatio-temporal high-resolution vortical flows can be obtained from only two coarse snapshot data.
It should be note that linear interpolation in time cannot capture advective physics.
They demonstrated the model capability with turbulent channel flows and reported that the flow field can be quantitatively reconstructed, achieving 0.04\% data compression.
Arora and Shrivastava~\cite{AS2022} have recently combined this super-resolution/inbetweening idea with physics-informed neural network~\cite{raissi2019physics} to improve the reconstruction accuracy and demonstrated it with an example of a mixed-variable elastodynamics system.

Furthermore, supervised super-resolution reconstruction can be used to examine how machine learning extracts the relationship between small and large-scale vortical structures.
Kim and Lee~\cite{KL2020} considered a CNN-based estimation of the high-resolution heat flux field in a turbulent channel flow from poorly-resolved wall-shear stresses and pressure.
They revealed that the CNN model focuses on the relationship between vortical structures and pressure distribution in channel turbulence to estimate the local heat flux from the wall-shear stress.
Morimoto et al.~\cite{morimoto2022generalization} has recently examined the effect of inter- and extrapolation with machine-learning-based super-resolution reconstruction with respect to flow parameters.
They considered two-staggered cylinder wakes whose flow dynamics are characterized based on the diameters and the distance between two cylinders.
They found that the supervised CNN-based model can quantitatively reconstruct a vortical flow even for untrained parameter cases by preparing flow field data based on the information of lift coefficient spectrum.

Supervised super-resolution techniques have also been applied to larger-scale meteorological flows~\cite{onishi2019super,yasuda2022super}.
Onishi et al.~\cite{onishi2019super} proposed a CNN-based model for super-resolution analysis of temperature fields in urban environment.
The proposed model provides a high-resolution temperature field at reduced computational time than the corresponding high-fidelity simulation, suggesting the potential use of machine-learning models as a surrogate for large-scale numerical simulations. 
To improve the model performance, Yasuda et al.~\cite{yasuda2022super} extended their model by incorporating skip connection~\cite{he2016deep} and channel attention~\cite{hu2018squeeze}.
While skip connection~\cite{he2016deep} helps stabilize the learning process of deep CNNs, channel attention~\cite{hu2018squeeze} can discover the crucial and irrelevant spatial regions of fluid flow regressions.
The model trained with temperature fields in one city (Tokyo) provides quantitative reconstruction for test temperature data for another city with similar climate (Osaka).
They also observed that including building height information as a part of the input of the machine-learning model is important for successful temperature reconstruction.

% Kuehn et al.~\cite{kuehn2022neural} has recently applied the DSC/MS model~\cite{FFT2019a} for super-resolution reconstruction of coastal sea states near Biarritz in France.
% They presented that the DSC/MS model is able to provide quantitative reconstruction with 50 times faster computational speed, compared to a reference high-resolution simulation.
% These studies shed a light on the capabilities of supervised-learning-based super-resolution analyses for practical applications in fluid dynamics.

In addition to the aforementioned studies with numerical data, applications to experiments have also been considered~\cite{MLPIV2023}.
For such cases, the effects of noise in the input data must be carefully considered.
Deng et al.~\cite{DHLK2019} developed a machine-learning model to super-resolve PIV measurements.
For training the model based on CNN, a pair of high-resolution experimental velocity data collected by PIV with cross-correlation method and downsampled low-resolution data is used.
The model was tested for turbulent flows around a single cylinder and two cylinders.
For more complex turbulent flows, Wang et al.~\cite{wang2020predicting} proposed a super-resolution neural
network for two-dimensional PIV (PIV2DSR) based on CNNs.
Once they trained the model with velocity fields of turbulent channel flow at $Re_\tau = 1000$ obtained by direct numerical simulation (DNS), the model is assessed with not only numerical channel flow field data at a much higher Reynolds number of 5200 but also real experimental PIV data for a turbulent boundary layer at $Re_\tau = 2200$.

%% FIGURE 1 %%%%%%%%%%%%%%%%%%%%%%%%%%%%%%%%%%%%%%%%%%%
\begin{figure*}[]
    \centering
		%\hspace{-30mm}
		\includegraphics[width=0.95\textwidth]{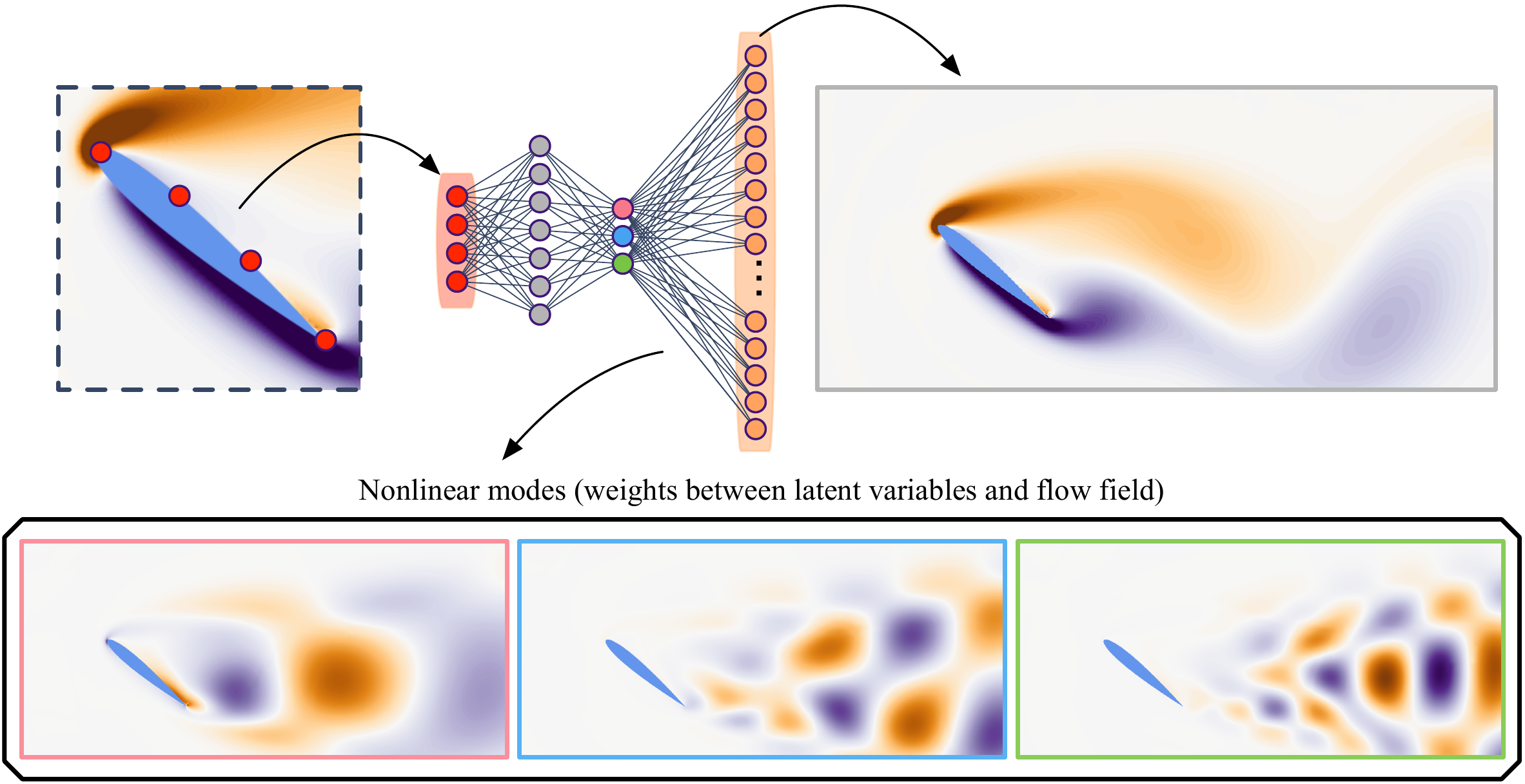}
		\caption{Extraction of nonlinear modes~\cite{MFF2019} from shallow decoder~\cite{erichson2020shallow} in super-resolution reconstruction for an example of two-dimensional incompressible flow (vorticity field) over a NACA0012 airfoil ($Re=100$ and $\alpha=40$ deg).}
		\label{fig_nonmodes}
\end{figure*}
%%%%%%%%%%%%%%%%%%%%%%%%%%%%%%%%%%%%%%%%%%%%%%%%%%%%%%

For the preparation of training data in these experimental studies, cross-correlation methods~\cite{adrian2005twenty} are generally used to obtain velocity fields from particle images.
Instead of giving a velocity field from the correlation method, one may consider providing a particle image directly into a model to obtain a higher-resolution flow field.
Cai et al.~\cite{CZXG2019} used FlowNetS~\cite{deconv2015} to estimate velocity fields of cylinder wake, backward-facing step flow, and isotropic turbulence from synthetic particle images.
They exhibited that a machine-learning model provides higher-resolution flow field data than the conventional PIV.
The proposed method was also tested with experimental particle images of a turbulent boundary layer.
Reconstructed flows based on machine learning may capture phenomena that cannot be observed with conventional techniques.
This FlowNetS-based method has recently been commercialized as AI-PIV~\cite{majewski2020developing}.
The super-resolution approach with particle images has also been applied to a wake around bluff bodies to remove the influence of reflection and halation in PIV measurements~\cite{morimoto2021experimental}.

Alternatively, a set of sparse sensor measurements can be considered as input to machine-learning models instead of the low-resolution flow data.
For instance, Erichson et al.~\cite{erichson2020shallow} used a fully-connected model to reconstruct a global flow field from local sensors.
The model was applied to geophysical flow and forced isotropic turbulence.
Their fully-connected model is a shallow decoder -- the model that incorporates a dimension compression to nonlinearly extract key features from sensors, after which the whole field is recovered from these latent representations of the input sensors, as illustrated in figure~\ref{fig_nonmodes}.
By visualizing the weight distribution between the latent space representation and the whole field, the shallow decoder provides nonlinear modes that represent the contribution of each latent variable for super-resolution reconstruction, which are analogous to those captured by nonlinear autoencoders~\cite{MFF2019,FT2022b,FNF2020,ricardo_VAE2021,FHNMF2020,linot2020deep,FT2023}.

\begin{figure*}[]
    \centering
		%\hspace{-30mm}
		\includegraphics[width=0.95\textwidth]{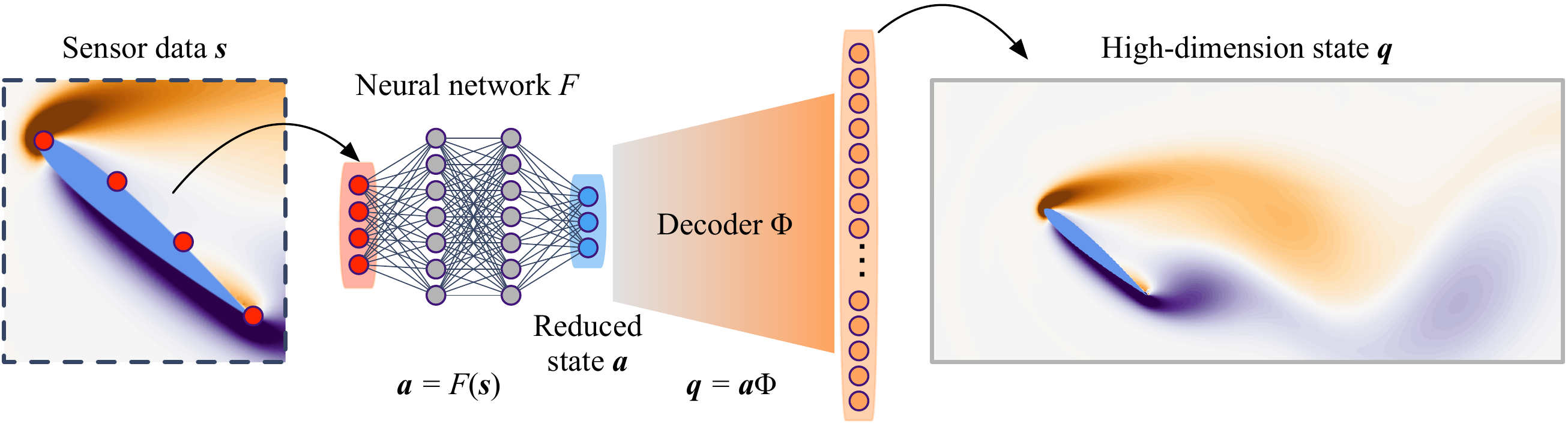}
		\caption{Reduced-order modeling-assisted super-resolution reconstruction~\cite{NG2020,dubois2022machine}.}
		\label{fig_NG}
\end{figure*}

As mentioned above, fully-connected network-based reconstruction is prohibitively expensive for global flow field reconstruction due to the very large number of parameters in the network~\cite{wu2020comprehensive}.
To address this issue, there are also some efforts to estimate low-order representations such as coefficients obtained through proper orthogonal decomposition (POD) from sparse sensor measurements~\cite{carter2021data,giannopoulos2020data,MFRFT2020}.
For instance, Nair and Goza~\cite{NG2020} proposed a fully-connected model-based estimator of POD coefficients and applied it to a laminar wake around a flat plate.
Their fully-connected model takes vorticity sensors on the airfoil surface and then outputs POD coefficients, as illustrated in figure~\ref{fig_NG}.
They considered wakes with two different angles of attacks, and reported that the neural-network model outperforms conventional linear techniques such as Gappy POD~\cite{ES1995,BDW2004} and linear stochastic estimation~\cite{adrian1988stochastic}.
Similarly, Manohar et al.~\cite{manohar2022sparse} has also recently performed a fully-connected model and POD-based sparse reconstruction for wake interactions of two cylinders.
Their model considers the time history of sensor measurements with long short-term memory (LSTM)~\cite{HS1997}, achieving more robustness against noisy inputs compared to a regular MLP model.
These reduced-order strategies in machine-learning-based vortical flow reconstruction are summarized in Dubois et al.~\cite{dubois2022machine}.
With flow examples of two- and three-dimensional cylinder wakes and a spatial mixing layer, they discussed pros and cons of a variety of techniques such as POD~\cite{Lumely1967,Holmes,TBDRCMSGTU2017}, regular autoencoder~\cite{HS2006}, variational autoencoder~\cite{rezende2014stochastic}, linear/nonlinear fully-connected networks, support vector machine~\cite{smola2004tutorial}, gradient boosting~\cite{friedman2001greedy}, and library-based reconstruction~\cite{BPK2016a,callaham2019robust}.
% Although the strategy with reduced-complexity techniques allows us to estimate a flow field by combining the estimated coefficients and the modes with less computational cost, users have to choose the number of modes depending on flows of interest.

%% FIGURE 1 %%%%%%%%%%%%%%%%%%%%%%%%%%%%%%%%%%%%%%%%%%%
%%%%%%%%%%%%%%%%%%%%%%%%%%%%%%%%%%%%%%%%%%%%%%%%%%%%%%

From the aspect of reducing the number of parameters inside machine-learning models, a combination of a fully-connected model and CNNs has also been leveraged to overcome the limitation of fully-connected networks.
Morimoto et al.~\cite{MFMVF2022} considered a combination of multi-layer perceptron (MLP) and CNN (called MLP-CNN-based estimator) to estimate vortical flows around urban structures and temperature data (DayMET) across North America from sparse sensors.
The sensor inputs are first given into the part of a fully-connected model and the model extracts the features from the input sensors.
The feature vectors extracted from it are then given to the convolutional layers.
Compared to solely using fully-connected layers, the computational cost can be significantly reduced while maintaining the reconstruction accuracy.
A similar MLP-CNN model was also considered by Zhong et al.~\cite{ZFAT2022,ZFAT2023} for a vortex-airfoil gust interaction problem.
The model estimates a two-dimensional vorticity field from pressure sensor measurements on an airfoil surface.
They reported that transfer learning~\cite{pan2009survey,lee2022predicting} can help in reducing the required amount of training data, while recurrent neural network (long short-term memory, LSTM~\cite{HS1997}) also improves the reconstruction performance of complex transient wake problems.

\subsection{Semisupervised- and unsupervised learning}
\label{sec:sul}

In addition to supervised-learning-based efforts, semisupervised- and unsupervised learning can be used in super-resolution analysis of fluid flows.
Semisupervised learning combines a small amount of labeled data with a large amount of unlabeled data, which can also be augmented with prior knowledge incorporated into the loss function.
Gao et al.~\cite{gao2021super} proposed a semisupervised CNN-based super-resolution analysis for fluid flows.
Through the investigation of a two-dimensional laminar flow and a cardiovascular flow, they showed that the constraints based on the conservation laws and boundary conditions enable successful super-resolution reconstruction without high-resolution labeling.
These physics-law-based augmentations inspired by physics-informed neural network (PINN)~\cite{lagaris1998artificial,raissi2019physics,raissi2020hidden} achieve accurate reconstruction while reducing the required amount of training data~\cite{yousif2021high,yousif2022deep}.

There are also a couple of studies on semisupervised super resolution.
Bode et al.~\cite{bode2021using} proposed the physics-informed enhanced super-resolution generative adversarial network (PIESRGAN) for applications to subgrid-scale modeling of LES.
To incorporate a physics-based loss function, they used the following cost function~$\cal E$ for training,
\begin{align}
    {\cal E} = {\cal E}_{\rm adv} + \beta_{\rm reg} {\cal E}_{\rm reg} + \beta_{\rm grad} {\cal E}_{\rm grad} + \beta_{\rm cont} {\cal E}_{\rm cont},
\end{align}
where $\beta_{\rm reg}$, $\beta_{\rm grad}$, and $\beta_{\rm cont}$ are weighting coefficients for the different loss term contributions.
The first loss term ${\cal E}_{\rm adv}$ corresponds to a regular adversarial loss used in GAN-based models, introduced in equation~\ref{eq_GAN}~\cite{wang2018esrgan}.
The second term ${\cal E}_{\rm reg}$ is a regular supervised loss function, which is equivalent to equation~\ref{eq_super}.
The PIESRGAN also includes the gradient loss ${\cal E}_{\rm grad}$ defined as the $L_2$ error norm of the gradient of state variables~\cite{bode2019deep}.
Weighting the gradient of the flow field promotes a smooth and physically-plausible reconstruction~\cite{HFMF2020a,HFMF2020b}.
They also considered ${\cal E}_{\rm cont}$, the divergence-free error for incompressible flow.
Similarly, a combination of physics-based loss and U-Net (figure~\ref{fig_Unet}) was proposed by Esmaeilzadeh et al.~\cite{esmaeilzadeh2020meshfreeflownet} as MeshfreeFlowNet and was applied for the Rayleigh-B\'{e}nard instability problem.
Due to the U-Net-based augmentation, the training for MeshfreeFlowNet takes only less than 4 minutes with 128 GPUs while achieving quantitative reconstruction.
To improve the generalizability of MeshfreeFlowNet~\cite{esmaeilzadeh2020meshfreeflownet} for a wide variety of problems, Wang et al.~\cite{TransFlowNet2022} have recently proposed TransFlowNet which weakens the constraint of initial and boundary conditions compared to MeshfreeFlowNet.
TransFlowNet was tested with examples of shallow water equation and Rayleigh-B\'{e}nard convection.
The model provides better reconstruction than the original MeshfreeFlowNet, although the instability of training process is also observed due to the complexity of model.

%% FIGURE 1 %%%%%%%%%%%%%%%%%%%%%%%%%%%%%%%%%%%%%%%%%%%
\begin{figure*}
    \centering
		%\hspace{-30mm}
		\includegraphics[width=1\textwidth]{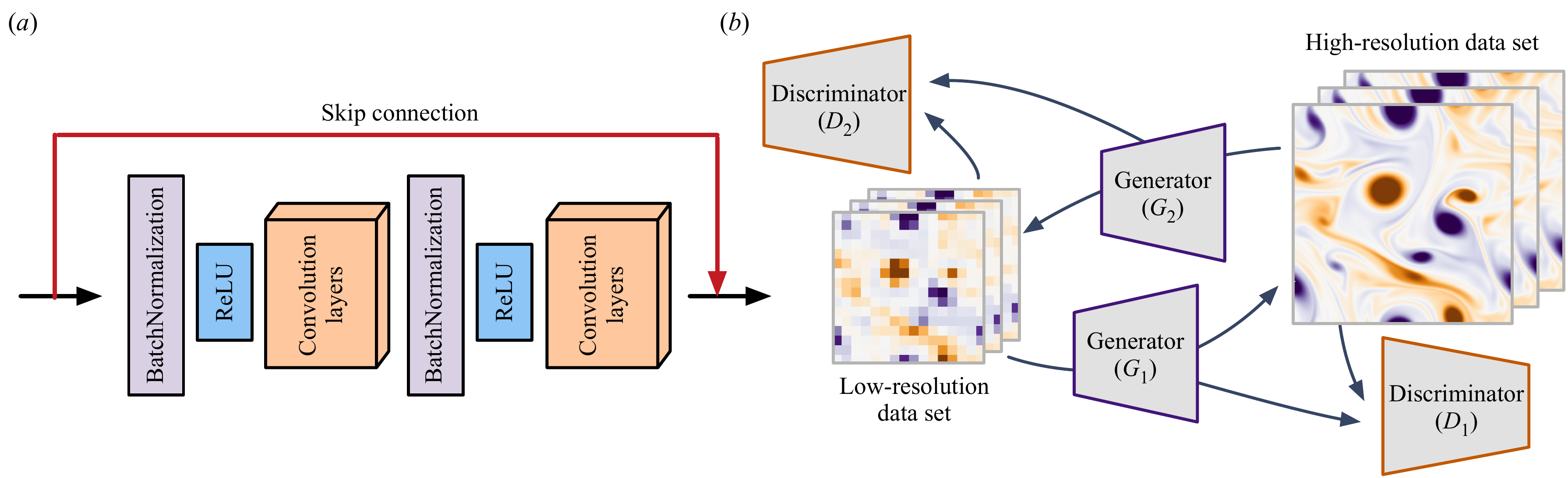}
		\caption{$(a)$ ResBlock~\cite{he2016deep} comprised of BatchNormalization layer, ReLU activation, and convolutional layer.
		$(b)$ Cycle GAN (cGAN)~\cite{zhu2017unpaired,MTK2023}.}
		\label{fig_RB}
\end{figure*}
%%%%%%%%%%%%%%%%%%%%%%%%%%%%%%%%%%%%%%%%%%%%%%%%%%%%%%

While incorporating the aforementioned physics loss can promote a physically-plausible super-resolution solution, we should be mindful of the fact that finding an appropriate balance between the weighting coefficients is challenging.
We can consider the use of optimization for finding an optimal set of coefficients, although it is computationally expensive~\cite{psaros2022meta}.
The influence of balancing between an adversarial error and a regular $L_2$ reconstruction error for sparse flow reconstruction is discussed in detail by Zhang et al.~\cite{zhang2022towards} for an example of a flow around building models.
Moreover, achieving stable convergence during training is also difficult with such complex loss functions.
To avoid this issue, additional machine-learning functions such as skip connection~\cite{he2016deep} and BatchNormalization~\cite{ioffe2015batch} can be leveraged.
In fact, the aforementioned models such as PIESRGAN~\cite{bode2019deep}, MeshfreeFlowNet~\cite{esmaeilzadeh2020meshfreeflownet}, and TransFlowNet~\cite{TransFlowNet2022} are composed of ResBlock~\cite{he2016deep} (illustrated in figure~\ref{fig_RB}$(a)$), which includes both BatchNormalization and skip connection, for stable and successful learning.

In contrast to supervised and semisupervised learning, unsupervised learning, which does not require labeled data sets, is also used for super-resolution analysis.
Kim et al.~\cite{kim2021unsupervised} proposed a cycle generative adversarial network (cGAN)-based framework for unsupervised super-resolution reconstruction of turbulent flows.
While a regular GAN is composed of one generator and one discriminator as presented in section~\ref{sec:GAN}, cGAN possesses two generators ($G_1$ and $G_2$) and two discriminators ($D_1$ and $D_2$), as illustrated in figure~\ref{fig_RB}$(b)$.
One generator $G_1$ attempts to reconstruct a high-resolution data ${\bm q}_{\rm HR}$ from a low-resolution flow field ${\bm q}_{\rm LR}$, while another generator $G_2$ provides low-resolution fields from the generated high-resolution flow data through $G_1$.
The discriminators $D_1$ and $D_2$ are trained to distinguish the real data from the generated data, as depicted in figure~\ref{fig_RB}$(b)$.
This operation allows the cGAN model to learn common features between low- and high-resolution data, that need not be paired~\cite{zhu2017unpaired}.
The proposed model can reconstruct a velocity field of turbulent channel flow from its low-resolution field.
They also demonstrated that the model trained with data from DNS can be applied to the LES data.

Following the study by Kim et al.~\cite{kim2021unsupervised}, the unsupervised GAN-based super resolution has recently been examined for a variety of flows.
Wurster et al.~\cite{wurster2021deep} proposed a hierarchical GAN to perform super resolution of fluid flows.
Analogous to SURFNet~\cite{obiols2021surfnet}, a hierarchical GAN is first trained with low-resolution data sets.
The model weights are then transferred to training with higher-resolution flow fields.
G{\"u}emes et al.~\cite{guemes2021coarse} combined a GAN-based super-resolution reconstruction and state estimation~\cite{BMT2001,CHBH2006,CCB2011,SH2017} from the wall sensor measurements of turbulent channel flow.
They first perform super-resolution reconstruction for wall-shear stresses and wall pressure. 
Another GAN model is then constructed to estimate wall-parallel velocity fields at several wall-normal locations from the super-resolved wall measurements.
The GAN models are able to provide reasonable agreement with the reference simulation data up to $y^+\approx 50$.
Yousif et al.~\cite{yousif2022super} extended a super-resolution GAN model by combining it with multi-scale CNN~\cite{FFT2019a} and applied it to a turbulent channel flow with large longitudinal ribs.
The reconstructed flow fields are shown to retain the temporal correlations and high-order spatial statistics.

Moreover, the use of a CNN-based GAN for three-dimensional super-resolution analysis was examined by Xu et al.~\cite{xu2020data} for computed tomography (CT) of turbulent jet combustor.
With an example of turbulent atmospheric flow, Hassanaly et al.~\cite{hassanaly2022adversarial} has comprehensively compared various models for super-resolution reconstruction, including a super-resolution GAN~\cite{ledig2017photo}, stochastic estimation, a deconvolution GAN~\cite{stengel2020adversarial}, and diversity-sensitive conditional GAN~\cite{yang2019diversity}.
Although GAN-based models have issues with stability during the learning process, these models hold potential for high-wavenumber reconstruction of turbulent flows.

\section{Case study: super-resolution reconstruction of turbulence}
\label{sec:results}

This section offers details of CNN-based super-resolution reconstruction for fluid flows through a case study.
As an example, we consider two-dimensional decaying isotropic turbulence, which serves as a canonical turbulent flow.
% With the wide use of CNNs for super-resolution analysis, we hope that the insights gained from the present case study will be useful for a wide variety of super-resolution models.
The flow field data to be studied is generated by a two-dimensional DNS~\cite{TNB2016}, which numerically solves the two-dimensional vorticity transport equation,
\begin{equation}
\dfrac{\partial\omega}{\partial t}+{\bm u}\cdot\nabla\omega=\dfrac{1}{Re_0}\nabla^2 \omega,
\label{eq_1}
\end{equation} 
where ${\bm u}=(u,v)$ and $\omega$ represent the velocity and vorticity fields, respectively.  
The computational domain is a biperiodic square with $L_x=L_y=1$.
The initial Reynolds numbers for training/validation and test data sets are respectively set to $Re_0\equiv u^*l_0^*/\nu=\{451,442\}$.
Here, $u^*$ is the characteristic velocity defined as the square root of the spatially averaged initial kinetic energy, {$l_0^*=[2{\overline{u^2}}(t_0)/{\overline{\omega^2}}(t_0)]^{1/2}$} is the initial integral length, and $\nu$ is the kinematic viscosity.
The numbers of computational grid points used by DNS are $N_x=N_y=512$.
For training the baseline networks, we use 1000 snapshots over an eddy turn-overtime of $t\in[2,6]$ with a time interval of $\Delta t=0.004$.
We consider a vorticity field $\omega$ as the variable of interest.

We note that our previous studies~\cite{FFT2019a,FFT2021b} on machine-learning-based super-resolution reconstruction was performed with two-dimensional decaying turbulence but at lower Reynolds numbers ($Re_0\approx 80$) with smaller numbers of the grid points ($N=128$).
The present case study examines how the model can be improved with regard to not only reconstruction accuracy but also a large amount of necessary training data at a higher Reynolds number.

%% FIGURE 1 %%%%%%%%%%%%%%%%%%%%%%%%%%%%%%%%%%%%%%%%%%%
\begin{figure*}[b]
    \centering
		%\hspace{-30mm}
		\includegraphics[width=1\textwidth]{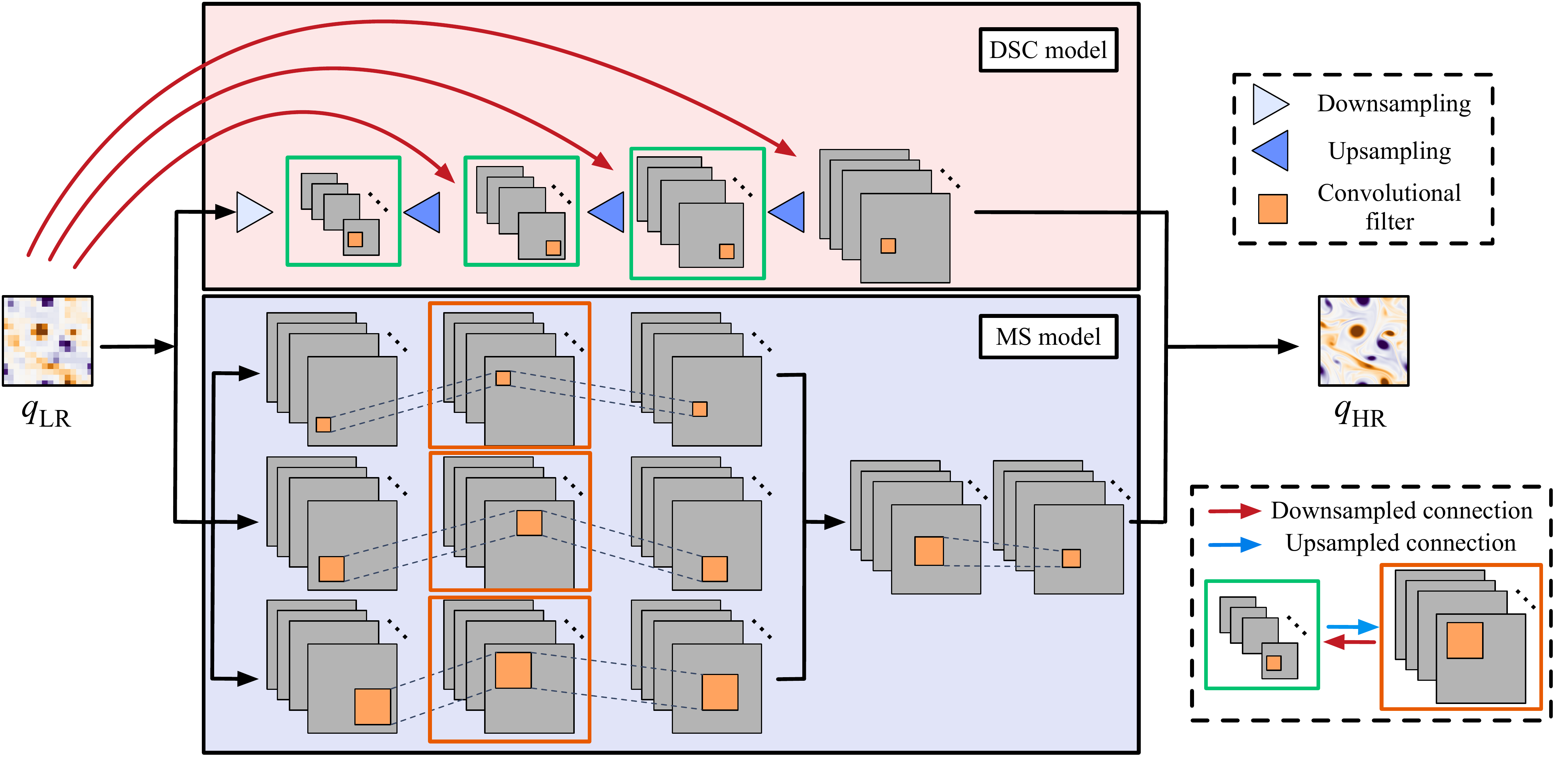}
		\caption{Interconnected DSC/MS model for super-resolution reconstruction of turbulent flows.}
		\label{fig2}
\end{figure*}
%%%%%%%%%%%%%%%%%%%%%%%%%%%%%%%%%%%%%%%%%%%%%%%%%%%%%%

For the present study, we consider super-resolution reconstruction with a regular CNN and the hybrid downsampled skip-connection/multi-scale (DSC/MS) model~\cite{FFT2019a}.
The design of the DSC/MS model is illustrated in figure~\ref{fig2}.
The red portion of the downsampled skip connection (DSC) model is composed of up-/downsampling operations and skip connection.
The up-/downsampling operations provide robustness against rotational and translational invariance.
The skip connection plays a crucial role in learning hierarchically the relationship between the high-resolution output and the low-resolution input, while providing numerical stability during the learning process of the CNN~\cite{he2016deep}.
The present model also incorporates the multi-scale model (MS) model~\cite{DQHG2018}, corresponding to the blue portion of figure~\ref{fig2}.
This part of the model performs filtering operations across three different sizes, capturing a range of spatial length scales in vortical flows.

To accurately reconstruct two-dimensional higher Reynolds number turbulent flow, we provide additional internal skip connections between the DSC model and MS model, as depicted in the green and orange boxes in figure~\ref{fig2}.
Each green box in the DSC model connects with each of the orange boxes in the MS model, hence nine connections are present.
With these interconnections, this interconnected DSC/MS model enables the intermediate input/output from both submodels to correlate with each other through the learning process.
Since coverage of spatial length scales increases with the Reynolds number, the interconnections are expected to be important in learning the relationship between small and large vortical elements by the model.
For the activation function $\varphi$, this study uses the ReLU function~\cite{NH2010} to avoid vanishing gradients of weights during the training process.

Furthermore, we consider a physics-based loss function to examine its effects on machine-learning-based super-resolution reconstruction of turbulent vortical flows.
As discussed in section~\ref{sec:PILF}, the use of physics-inspired loss function may not only promote the physical validity of reconstruction but also reduce the amount of necessary training data in a semisupervised manner~\cite{gao2021super,yousif2021high,esmaeilzadeh2020meshfreeflownet,ren2022physics}.
Here, we use the nonlinear advection term and the linear viscous diffusion term in equation~\ref{eq_1} for the physics-based loss function.
The present cost function ${\cal E}$ is hence defined as
\begin{align}
    {\cal E}   &= {\cal E}_{\omega} + \beta_{\rm adv}{\cal E}_{\rm adv} + \beta_{\rm visc}{\cal E}_{\rm visc},~~~{\rm where}~\label{eq:err}\\
    {\cal E}_{\omega} &= ||\omega_{\rm DNS}-F(\omega_{\rm LR})||_2,\nonumber\\ 
    {\cal E}_{\rm adv} &= ||{\bm u}_{\rm DNS}\cdot\nabla\omega_{\rm DNS}-{\bm u}_{\rm ML}\cdot\nabla\omega_{\rm ML}||_2,\nonumber\\
    {\cal E}_{\rm visc} &= ||\nabla^2 \omega_{\rm DNS}-\nabla^2 \omega_{\rm ML}||_2,\nonumber
\end{align}
in which $\omega_{\rm DNS}$ and $\omega_{\rm LR}$, respectively, represent the reference (high-resolution) DNS field and the low-resolution input flow field.
The coefficients $\beta_{\rm adv}$ and $\beta_{\rm visc}$ determine the balance of the terms in the loss function.
The terms $(\cdot)_{\rm ML}$ inside ${\cal E}_{\rm adv}$ and ${\cal E}_{\rm visc}$ are computed with the super-resolved vorticity field $F(\omega_{\rm LR})$.

%% FIGURE 1 %%%%%%%%%%%%%%%%%%%%%%%%%%%%%%%%%%%%%%%%%%%
\begin{figure*}[b]
    \centering
		%\hspace{-30mm}
		\includegraphics[width=0.8\textwidth]{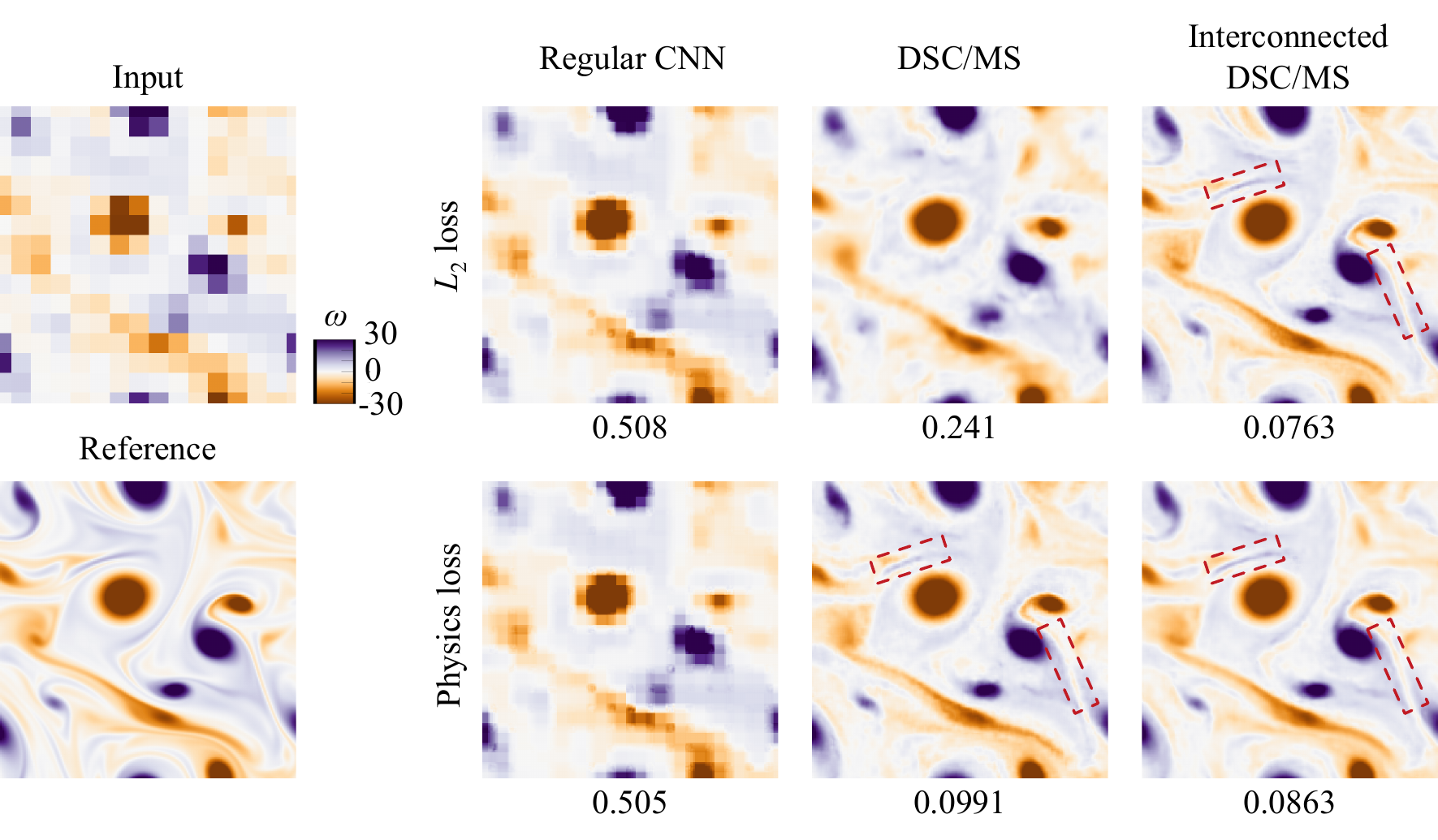}
		\caption{Super-resolution reconstruction of two-dimensional decaying homogeneous isotropic turbulence.
		The value underneath each vorticity contour plot presents the $L_2$ norm of reconstruction error $\epsilon$.
		}
		\label{fig3}
\end{figure*}
%%%%%%%%%%%%%%%%%%%%%%%%%%%%%%%%%%%%%%%%%%%%%%%%%%%%%%
%% FIGURE 1 %%%%%%%%%%%%%%%%%%%%%%%%%%%%%%%%%%%%%%%%%%%
\begin{figure*}[t]
    \centering
		%\hspace{-30mm}
		\includegraphics[width=0.83\textwidth]{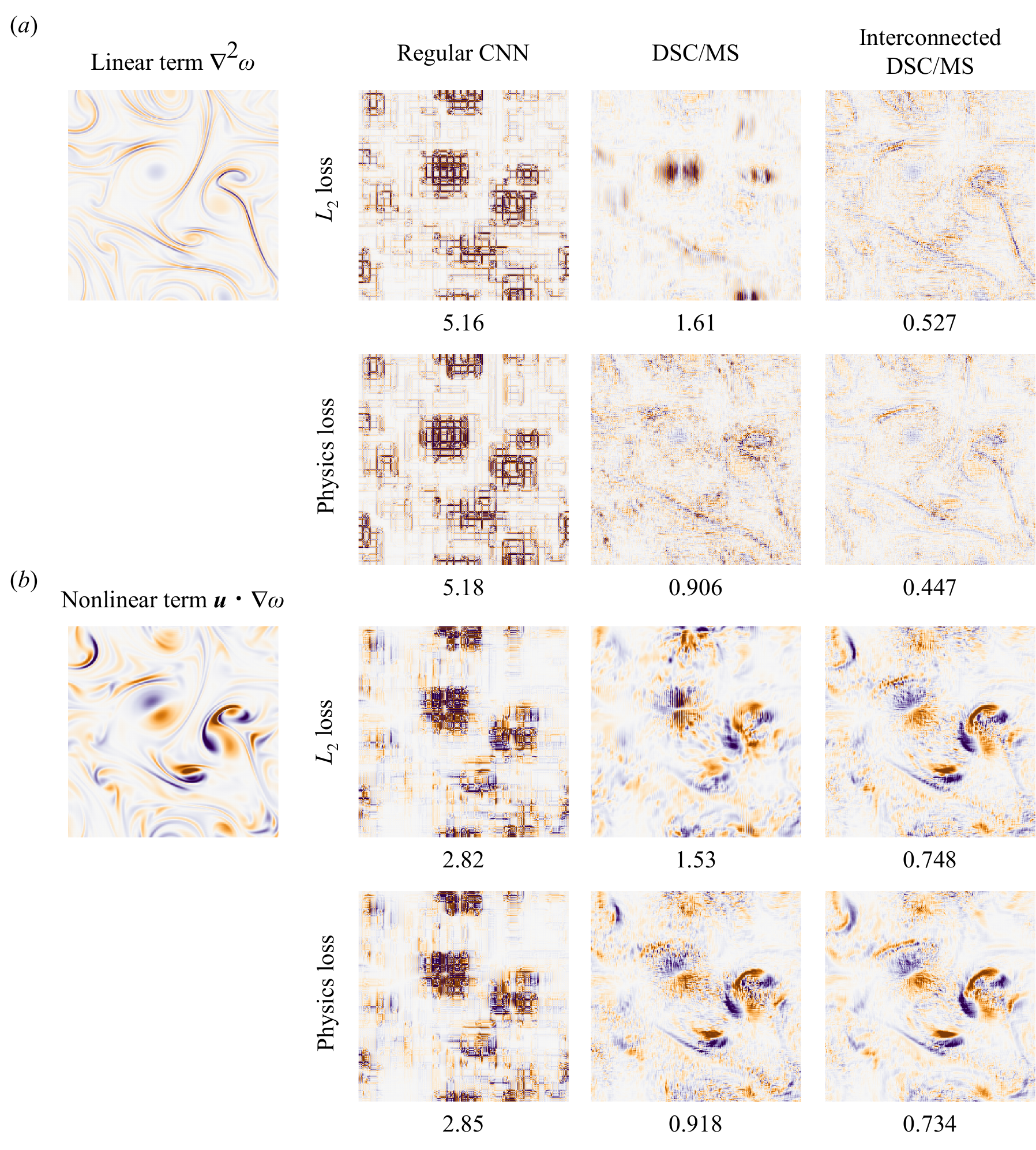}
		\caption{$(a)$ The linear term $\nabla^2\omega$ and $(b)$ the nonlinear term ${\bm u}\cdot\nabla\omega$, computed from the reconstructed flow fields for each machine-learned model.
		The value underneath each contour presents the $L_2$ norm error $\epsilon$.
		Shown results are from the same case as the vorticity snapshots presented in figure~\ref{fig3}.
		}
		\label{fig4}
\end{figure*}
%%%%%%%%%%%%%%%%%%%%%%%%%%%%%%%%%%%%%%%%%%%%%%%%%%%%%%

In what follows, we assess six different machine-learning models:
\begin{enumerate}
\item {CNN-$L_2$}: a regular CNN model with ${\cal E} = {\cal E}_{\omega}$,
\item {CNN-$L_{\rm phys}$}: a regular CNN model with ${\cal E} = {\cal E}_{\omega} + \beta_{\rm adv}{\cal E}_{\rm adv} + \beta_{\rm visc}{\cal E}_{\rm visc}$,
\item {DSC/MS-$L_2$}: the original DSC/MS model~\cite{FFT2019a} with ${\cal E} = {\cal E}_{\omega}$,
\item {DSC/MS-$L_{\rm phys}$}: the original DSC/MS model with ${\cal E} = {\cal E}_{\omega} + \beta_{\rm adv}{\cal E}_{\rm adv} + \beta_{\rm visc}{\cal E}_{\rm visc}$,
\item {IDSC/MS-$L_2$}: the interconnected DSC/MS model with ${\cal E} = {\cal E}_{\omega}$,
\item {IDSC/MS-$L_{\rm phys}$}: the interconnected DSC/MS model with ${\cal E} = {\cal E}_{\omega} + \beta_{\rm adv}{\cal E}_{\rm adv} + \beta_{\rm visc}{\cal E}_{\rm visc}$.
\end{enumerate}
These six machine-learning models are tasked to reconstruct the high-resolution vortical flow field of size $512^2$ from the corresponding low-resolution data of size $16^2$, generated by average-pooling operations~\cite{FFT2019a}.  
We set $\beta_{\rm adv} = \beta_{\rm visc} = 0.1$ to the balance of the order for each term.

Let us consider the reconstructed vorticity fields from machine-learing-based super-resolution approaches in figure~\ref{fig3}.
The large-scale vortices can be reconstructed with the regular CNN models.
However, the reconstructed fields are pixelized around rotation and shear-dominated structures, which was also observed with a regular CNN-based super-resolution reconstruction in our previous study~\cite{FFT2019a}.
The $L_2$ norm error, $\epsilon = ||{f_{\rm DNS}}-{f_{\rm ML}}||_2/||{f_{\rm DNS}}||_2$, is found to be larger than 0.5 with the regular CNNs.
The DSC/MS model with the $L_2$-based optimization provides a better and clear reconstruction for large vortical structures with an $L_2$ norm error of 0.241.
This indicates that embedding the physics-inspired DSC functions and the MS filters enables accurate reconstruction of vortical flows.

While the DSC/MS model achieves a qualitative reconstruction for vortical structures, finer scales of shear layers that appear around large rotational elements cannot be recovered well.
The reconstruction over these scales that emerge in higher Reynolds number flows can be improved by introducing either the physics-based loss function or the interconnection inside the DSC/MS model, as presented in figure~\ref{fig3}.
With DSC/MS-$L_{\rm phys}$, IDSC/MS-$L_2$, and IDSC/MS-$L_{\rm phys}$, these shear layers are more accurately reconstructed compared to the reconstruction with the regular model, as highlighted by the red boxes in figure~\ref{fig3}.
Hence, both the physics-inspired optimization and model design greatly assist in the reconstruction of higher Reynolds number flows.
Note that the difference between the interconnection-based model enhancement and using the physics-based loss function is in their robustness against noisy low-resolution input, as it will be discussed later.

We here examine each term in the physics-loss function; namely the linear term $\nabla^2\omega$ and the nonlinear term ${\bm u}\cdot\nabla\omega$ of the present super-resolution reconstruction, as shown in figure~\ref{fig4}.
These results are from the same case as the vorticity snapshots presented in figure~\ref{fig3}.
Examination of these terms is a strict test since higher-order derivations can amplify errors greatly for high wavenumbers.
Let us first focus on the estimated linear viscous diffusion term visualized in figure~\ref{fig4}$(a)$.
The regular CNN completely fails to estimate $\nabla^2\omega$, as evident from the pixelized vorticity reconstruction in figure~\ref{fig3}.
Using the DSC/MS model with the regular $L_2$ optimization, the linear term field also exhibits erroneous profiles comprised of pairwise structures that cannot be observed in the reference field.
These derivative-based assessments are again very sensitive and also affected by the reconstruction of surrounding local structures.
As expected, estimation is improved by including the physics-based term in the loss function (DSC/MS-$L_{\rm phys}$).
The accuracy of $\nabla^2\omega$ can be further enhanced by using the interconnected DSC/MS models, presenting fine-scale structures in the high-order derivation field.
This indicates that adding the interconnections inside the machine-learning model enables physically-compatible super-resolution reconstruction of turbulent flows in addition to the physics-loss-based optimization.

The estimation of the nonlinear term is shown in figure~\ref{fig4}$(b)$.
The whole trend in reconstruction is analogous to that for the linear term, hence the interconnected DSC/MS models well reconstruct the nonlinear term fields compared to the reference field.
The $L_2$ errors for the nonlinear term with DSC/MS-$L_{\rm phys}$, IDSC/MS-$L_2$, and IDSC/MS-$L_{\rm phys}$ are higher than that of the linear term.
This suggests that the estimation for the nonlinear term is more difficult than the linear term.

\begin{figure*}
    \centering
		%\hspace{-30mm}
		\includegraphics[width=0.85\textwidth]{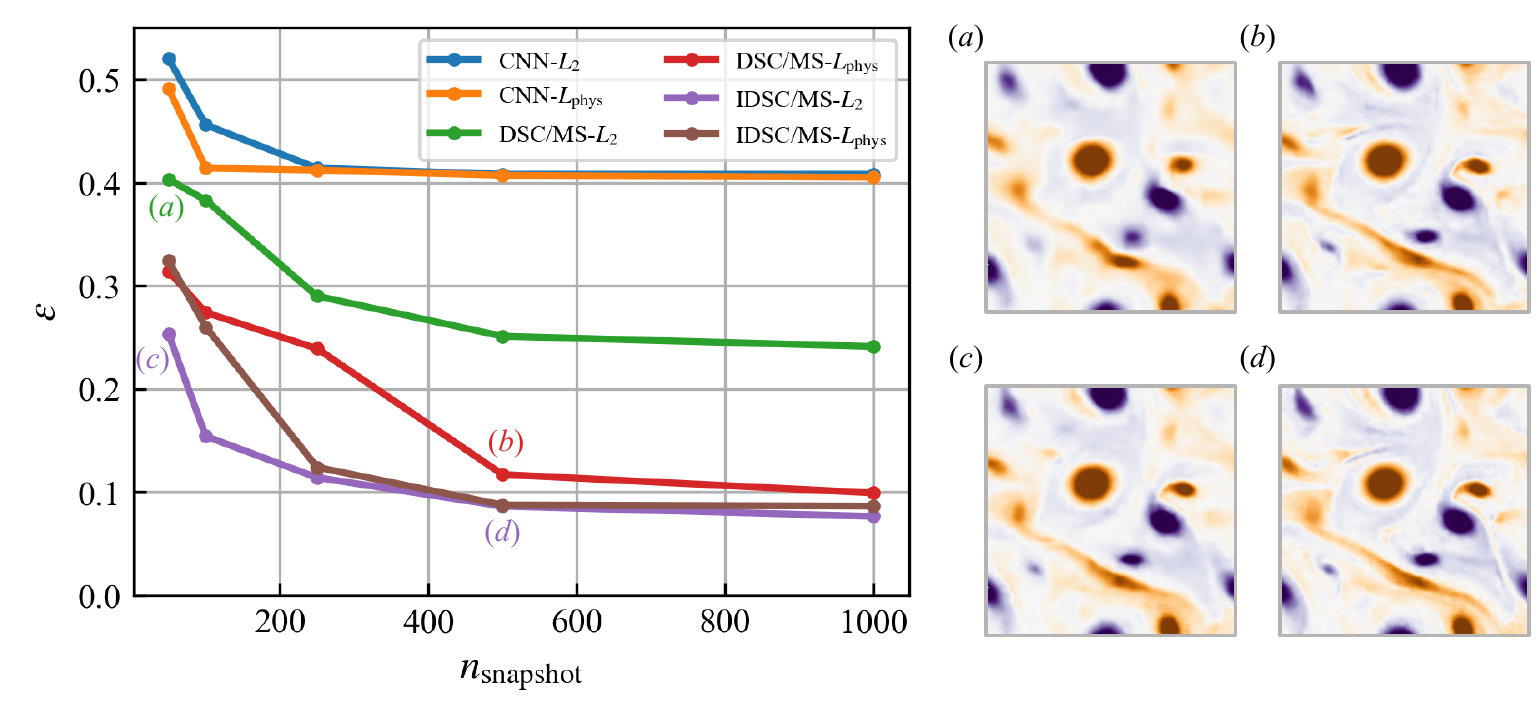}
		\caption{Dependence of the reconstruction accuracy on the number of the training snapshots.}
		\label{fig6}
\end{figure*}

We also investigate the dependence of the reconstruction error on the number of the training snapshots~$n_{\rm snapshot}$.
For all models, the error decreases as $n_{\rm snapshot}$ increases, as shown in figure~\ref{fig6}.
Both the interconnection and the physics-based loss enable a qualitative reconstruction with a reduced number of training snapshots.
The observation that a physics-inspired optimization reduces the required amount of training data has also been reported in previous studies~\cite{gao2021super,raissi2020hidden}.
The interconnected DSC/MS model reconstructs fine vortical structures even with only $n_{\rm snapshot} = 50$ (figure~\ref{fig6}$(c)$), while the original DSC/MS model provides only large-scale structures, as shown in figure~\ref{fig6}$(a)$.
This suggests that the present machine-learning model efficiently captures a nonlinear relationship between the under-resolved input and the high-resolution vortical flow from a small amount of training data by capitalizing on the interconnected skip connections.

%% FIGURE 1 %%%%%%%%%%%%%%%%%%%%%%%%%%%%%%%%%%%%%%%%%%%
%%%%%%%%%%%%%%%%%%%%%%%%%%%%%%%%%%%%%%%%%%%%%%%%%%%%%%
%% FIGURE 1 %%%%%%%%%%%%%%%%%%%%%%%%%%%%%%%%%%%%%%%%%%%
\begin{figure*}
    \centering
		%\hspace{-30mm}
		\includegraphics[width=1\textwidth]{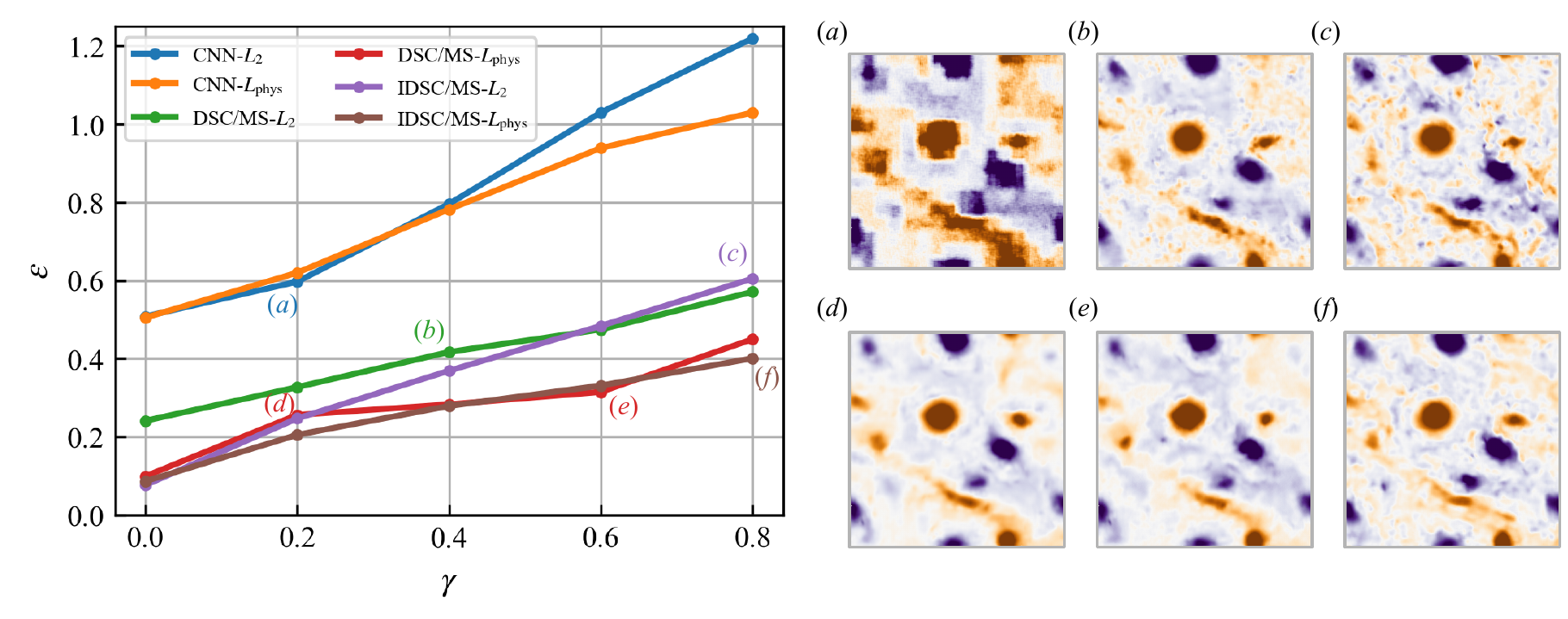}
		\caption{Dependence of the reconstruction accuracy on the magnitude of noisy input.}
		\label{fig7}
\end{figure*}
%%%%%%%%%%%%%%%%%%%%%%%%%%%%%%%%%%%%%%%%%%%%%%%%%%%%%%

The use of the physics-based loss function can lead to robustness against noisy inputs.
Here, let us examine the influence of noise on super-resolution reconstruction.
We add the Gaussian noise ${\bm n}$ to the low-resolution input ${\omega}_{\rm LR}$, and assess the reconstruction $L_2$ error $\epsilon = ||\omega_{\rm HR} - F(\omega_{\rm LR} + {\bm n})||_2/||\omega_{\rm HR}||_2$, where the magnitude of the noise is given as $\gamma = ||{\bm n}||/||\omega||$. 
Here, the models trained with 1000 snapshots are used.
The relationship between the error and the noise magnitude is shown in figure~\ref{fig7}.
For all cases, the error increases with increasing magnitude $\gamma$.
The reconstructed flow fields generally reveal the large-scale vortices, while the finer scales are affected by the noisy input.
Especially for $\gamma>0.3$, the DSC/MS models with the physics-based loss function are observed to be more robust than models trained with the simple $L_2$ error optimization.
Hence, it can be argued that physics-based loss function helps in devising robust models against noisy measurements.

\section{Extensions}
\label{sec:outlook}

In the above sections, we surveyed various machine-learning-based super-resolution approaches and their applications to vortical flows.
Here, we discuss extensions of machine-learning-based super-resolution analysis beyond their basic applications.

\subsection{Changing input variable setups}

%% FIGURE 1 %%%%%%%%%%%%%%%%%%%%%%%%%%%%%%%%%%%%%%%%%%%
\begin{figure*}[]
    \centering
		%\hspace{-30mm}
		\includegraphics[width=1\textwidth]{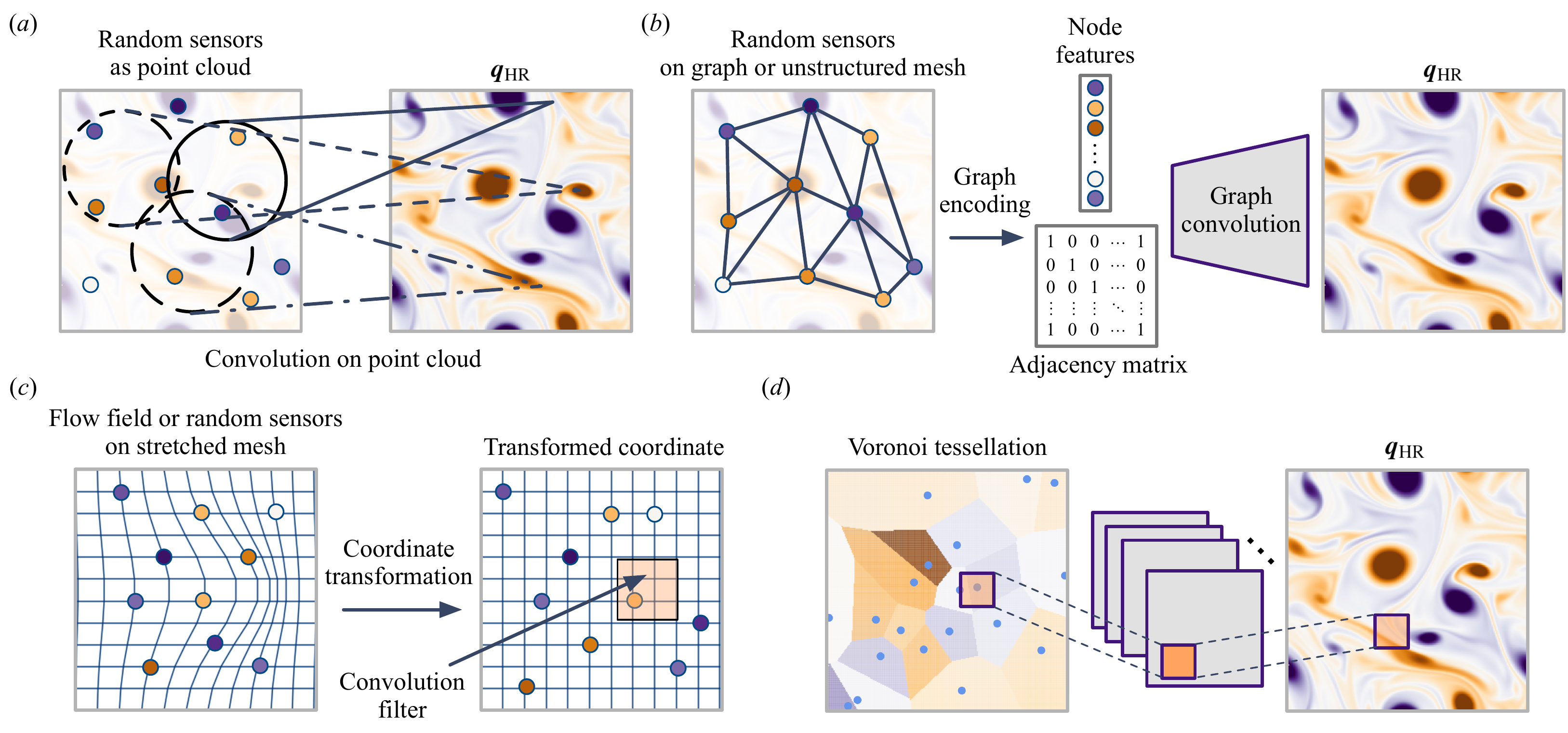}
		\caption{Applications of machine-learning-based super-resolution analysis for moving sensor and unstructured grid conditions. $(a)$ Convolution on point clouds~\cite{kashefi2021point,qi2017pointnet}.
		$(b)$ Graph neural network~\cite{liu2022fluid,gruber2022comparison}.
		$(c)$ Coordinate transformation~\cite{GSW2020}.
		$(d)$ Voronoi tessellation-based projection~\cite{fukami2021global}.
		}
		\label{fig8}
\end{figure*}
%%%%%%%%%%%%%%%%%%%%%%%%%%%%%%%%%%%%%%%%%%%%%%%%%%%%%%

When a machine-learning model is trained, the size of the input and output variables or more specifically the setup of the input and output variables is fixed.
If the setup is changed, the machine-learning model generally needs to be completely retrained, which is a heavy burden.
This issue is in fact a limitation of many machine-learning models, and machine-learning-based super-resolution models are no exceptions.
If the number of pixels or their locations is different from that used in the training process, the trained model cannot be used without retraining the model with the input variable size changed.  
Preprocessing the different-size input data with interpolation may work, but care should be taken since such an approach generally loses information. 
Unstructured grid and randomly sampled data also require some care since standard CNN-based models may not be appropriate.

There are several approaches to address these challenges.  
For instance, PointNet~\cite{qi2017pointnet} is able to handle unorganized and sparse data with a point cloud.  
Although this was originally used for image classification and segmentation tasks, Kashefi et al.~\cite{kashefi2021point} has recently applied it to fluid flows.  
In their formulation, sensors on the grid can be directly treated and a model can learn the relationship between the sensors and outputs, as illustrated in figure~\ref{fig8}$(a)$.

To handle spatially irregular sensor arrangements, graph neural network (GNN)~\cite{wu2020comprehensive} can be considered.  
GNN is able to perform a convolutional operation on unstructured mesh data, which is similar to that inside of CNNs.
% GNN-based approaches have recently been applied for reduced-order modeling in fluid flows~\cite{liu2022fluid,gruber2022comparison}.
Such GNN-based methods can be applied for machine-learning-based super-resolution reconstruction by modifying the setup for data dimensions between input and output, as shown in figure~\ref{fig8}$(b)$.

Coordinate transformation can also be considered to simply use regular machine-learning models for vortical flows.  
PhyGeoNet~\cite{GSW2020} includes coordinate transformation from an irregular domain to a structured mesh space for fluid flow regression, allowing us to convolve on flow fields, as illustrated in figure~\ref{fig8}$(c)$.
Finding the appropriate coordinate transformation may be a challenge for complex flow field domain geometry.

We can also generalize super-resolution analysis by considering sensor measurements in the flow field as the input for machine-learning models to reconstruct the flow field.
For a fixed number of sensors with their positions unchanged, regular machine-learning models developed in image science can often be directly used.  
However, when sensors go online or offline changing the number of sensors and moving spatially over time, the machine-learning models cannot be applied without special care.   
Voronoi-tessellation-based CNN~\cite{fukami2021global} can handle an arbitrary number of moving sensors in a single model.  
In this formulation, sparse sensor measurements are projected onto grids generated from Voronoi tessellation, as illustrated in figure~\ref{fig8}$(d)$. 
The flow field discretized with Voronoi tessellation is then used as an input for CNN-based super-resolution reconstruction.  
This approach provides robust real-time super-resolution analysis for vortical flows.

\subsection{Super resolution for turbulent flow simulations}

With the ability to recover fine-scale flow structures from coarse images of the flow field, it is natural to ask whether machine-learning-based super-resolution analysis can be incorporated into numerical simulations to improve turbulent flow simulations.  
From a broader perspective, this question translates to whether super-resolution analysis can be implemented in a simulation of multi-scale physical phenomena to accurately reconstruct the subgrid-scale physics~\cite{pradhan2021variational,bode2021using}.

For super-resolution analysis to reconstruct a physically accurate high-resolution flow field, it is generally necessary that the low-resolution input data is accurate on its own coarse grid. 
If the coarse flow field input is provided by some turbulent flow simulation (e.g., large-eddy simulation, detached eddy simulation, and Reynolds-averaged Navier-Stokes simulation~\cite{Kajishima}), it is important that the coarse flow be accurate to begin with.
The super-resolved field would not be a physically accurate if the low-resolution flow field (input) is deviated from the true solution. 
Conservatively speaking, turbulent flow statistics may be predicted well with super resolution but highly-accurate reconstruction of each and every instantaneous flow would likely be a major difficulty, if not impossible~\cite{kochkov2021machine}.  
In other words, we should not expect that LES results (or those from other solvers with turbulence models) can be transformed to yield DNS results.

A worthy question to ask is whether super-resolution analysis can support the development of subgrid-scale models.  
This could be different from other turbulence modeling approaches that use applied regressions to directly determine the subgrid-scale models for turbulent flow simulations.  
Similar to an approximate deconvolution model~\cite{stolz2001approximate} which considers inverse mapping of spatial filters, super resolution could be used to augment the subgrid-scale models.
Furthermore, it remains to be seen whether super-resolution analysis can simultaneously nudge the low-resolution field and recover the subgrid-scale flow structures.  
Again, the success of such simultaneous corrections will likely require the low-resolution flow field to be fairly accurate on its own grid.  
Alternatively, GAN-based techniques may also provide interesting approaches to achieve super resolution for turbulence.

Ongoing research developments in super-resolution analysis of turbulent flows and data-driven turbulence models~\cite{LKT2016,DIX2019} may address the issues identified here in the coming years.
As super-resolution methods are extended and incorporated into turbulent flow analysis and simulations, it is important to ensure that the derived super-resolution method is generalizable over a range of Reynolds numbers and turbulent flow problems to confirm robust and reliable performance.  
This is critical if these techniques are to be implemented in general-use turbulent flow simulators.

{
\subsection{Applications to real-world problems}

Super-resolution analysis holds great potential for fluid dynamics as discussed above.
However, there still exists some challenges, especially toward applications to real-world problems.
This section discusses the current challenges and possible future directions of machine-learning-based fluid flow super resolution.

One of the major challenges of machine-learning-based super-resolution reconstruction for fluid flows is the necessity for a certain amount of training data. 
While unsupervised learning used for GANs and semisupervised learning assisted with physics-inspired loss functions introduced in the present survey can mitigate this issue, existing techniques still require learning the relationship between coarse data and high-resolution vortical flows from either unpaired or paired training data for successful reconstruction.
Since the majority of real-world problems do not have access to ground truth and only sparse and noisy measurements are available, one can consider the use of data assimilation~\cite{du2023state,di2020synchronization,di2018inferring} to improve super-resolution reconstruction by incorporating the latest observations with a short-range real-time forecast.

Yasuda and Onishi~\cite{yasuda2022spatio} has recently proposed a four-dimensional super-resolution data assimilation and demonstrated its performance with a two-dimensional periodic channel flow.
The proposed method considers the temporal evolution of a system from low-resolution simulations with the aid of an ocean model, while a trained machine-learning model is simultaneously used to perform data assimilation and super resolution.
Since there is a huge amount of historical weather and climate reanalysis data available, the unification of super resolution with data assimilation or pre-existing models would be an interesting research direction.

In addition, most of the existing studies focus on designing a reconstruction model for a particular flow problem, variable, or data shape.
From this aspect, it would be desired to simultaneously leverage a variety of multi-modal data such as point-wise measurements, image-based data, and online measurements such as LiDAR-type data.
Prediction of unavailable parameters from such sparse and noisy but available measurements may also become an interesting direction of super-resolution studies of fluid dynamics.
}

\section{Conclusions}
\label{sec:conc}

We provided a survey on machine-learning-based super-resolution reconstruction of vortical flows.
Several machine-learning approaches and the use of physics-based cost functions for super-resolution analysis were discussed.
We further performed case studies of super-resolution reconstruction of turbulent flows with convolutional neural network (CNN)-based methods.
We demonstrated that a super-resolution model with physics-based loss function or physics-inspired neural network structures can reconstruct vortical flows even with limited training data and noisy inputs.
We also discussed extensions and challenges of machine-learning-based super resolution for fluid flows from the aspects of changing input variable setups and applications for turbulent flow simulations.

The insights obtained through the present survey can be leveraged for a variety of machine-learning-based super-resolution models.
For instance, the use of multi-scale filters inside CNN can be generalized not only in supervised learning but also in unsupervised techniques~\cite{yousif2022physics}.
Physics-informed loss functions can also be extended for various machine-learning models.
Moreover, it may also be interesting to develop super-resolution models in wavespace to incorporate certain spectral properties.

We remind that studies surveyed in the present paper are generally based on clean training data.
Preparing high-quality input data is essential for successfully reconstructing turbulent flows.
However, it is necessary to assess the robustness and sensitivity of the models against noisy inputs~\cite{nakamura2022identifying}.
This point will be important as machine-learning-based super-resolution analyses become utilized in industrial applications~\cite{Fukamipump2022}.
Together with the accuracy of the models, quantifying uncertainties in machine-learning prediction is also required to assess their reliability and limitations.
{For these reasons, making computational and experimental fluid-flow databases~\cite{li2008public,wu2008direct,towne2022database} available is critically important to advance studies on data-driven analysis of vortical flows.}
We hope that this survey paper provides some guidance in advancing algorithms and applications of machine-learning-based super-resolution analysis for a variety of fundamental and industrial fluid flow problems.

\section*{Acknowledgements}
K. Fukami acknowledges the support from the UCLA-Amazon Science Hub for Humanity and Artificial Intelligence.
K. Fukagata acknowledges the support by JSPS KAKENHI (grant numbers 18H03758 and 21H05007).
K. Taira acknowledges the generous support from the US Air Force Office of Scientific Research (grant FA9550-21-1-0178) and the US Department of Defense Vannevar Bush Faculty Fellowship (grant N00014-22-1-2798).

\section*{Declarations}

\subsection*{Conflict of interest}
The authors declare that they have no conflict of interest.
 
\subsection*{Authors' contributions}
Ka.F, Ko.F, and KT designed research. 
Ka.F performed research and analyzed data. 
Ka.F and KT wrote the paper.
Ko.F and KT supervised. 
 
\subsection*{Availability of data and materials }
The data that support the findings of this study are available from the corresponding author upon reasonable request.

\bibliographystyle{unsrt}  
\bibliography{arxiv}

\begin{thebibliography}{100}

\bibitem{irani1991improving}
M.~Irani and S.~Peleg.
\newblock Improving resolution by image registration.
\newblock {\em CVGIP: Graphical models and image processing}, 53(3):231--239,
  1991.

\bibitem{salvador2016example}
J.~Salvador.
\newblock {\em Example-Based super resolution}.
\newblock Academic Press, 2016.

\bibitem{bannore2009iterative}
V.~Bannore.
\newblock {\em Iterative-interpolation super-resolution image reconstruction: a
  computationally efficient technique}, volume 195.
\newblock Springer Science \& Business Media, 2009.

\bibitem{keys1981cubic}
R.~Keys.
\newblock Cubic convolution interpolation for digital image processing.
\newblock {\em IEEE Trans. Acoust. Speech Signal Process.}, 29(6):1153--1160,
  1981.

\bibitem{vandewalle2006frequency}
P.~Vandewalle, S.~S{\"u}sstrunk, and M.~Vetterli.
\newblock A frequency domain approach to registration of aliased images with
  application to super-resolution.
\newblock {\em EURASIP J. Adv. Signal Process.}, 2006:1--14, 2006.

\bibitem{joshi2008psf}
N.~Joshi, R.~Szeliski, and D.~J. Kriegman.
\newblock {PSF} estimation using sharp edge prediction.
\newblock In {\em IEEE Conference on Computer Vision and Pattern Recognition},
  pages 1--8. IEEE, 2008.

\bibitem{lucas1981iterative}
B.~D. Lucas and T.~Kanade.
\newblock An iterative image registration technique with an application to
  stereo vision.
\newblock {\em Proceedings of the DARPA Image Understanding Workshop},
  81:674–679, 1981.

\bibitem{michaeli2013nonparametric}
T.~Michaeli and M.~Irani.
\newblock Nonparametric blind super-resolution.
\newblock In {\em Proceedings of the IEEE International Conference on Computer
  Vision}, pages 945--952, 2013.

\bibitem{glasner2009super}
D.~Glasner, S.~Bagon, and M.~Irani.
\newblock Super-resolution from a single image.
\newblock In {\em IEEE 12th international conference on computer vision}, pages
  349--356. IEEE, 2009.

\bibitem{zontak2013separating}
M.~Zontak, I.~Mosseri, and M.~Irani.
\newblock Separating signal from noise using patch recurrence across scales.
\newblock In {\em IEEE conference on computer vision and pattern recognition},
  pages 1195--1202, 2013.

\bibitem{shahar2011space}
O~Shahar, A~Faktor, and M~Irani.
\newblock Space-time super-resolution from a single video.
\newblock In {\em CVPR 2011}, pages 3353--3360. IEEE Computer Society, 2011.

\bibitem{freedman2011image}
G.~Freedman and R.~Fattal.
\newblock Image and video upscaling from local self-examples.
\newblock {\em ACM Trans. Graph.}, 30(2):1--11, 2011.

\bibitem{yang2013fast}
J.~Yang, Z.~Lin, and S.~Cohen.
\newblock Fast image super-resolution based on in-place example regression.
\newblock In {\em Proceedings of the IEEE conference on computer vision and
  pattern recognition}, pages 1059--1066, 2013.

\bibitem{baker2002limits}
S.~Baker and T.~Kanade.
\newblock Limits on super-resolution and how to break them.
\newblock {\em IEEE Trans. Pattern Anal. Mach. Intell.}, 24(9):1167--1183,
  2002.

\bibitem{park2003super}
S.~C. Park, M.~K. Park, and M.~G. Kang.
\newblock Super-resolution image reconstruction: a technical overview.
\newblock {\em IEEE Signal Process. Mag.}, 20(3):21--36, 2003.

\bibitem{roweis2000nonlinear}
S.~T. Roweis and L.~K. Saul.
\newblock Nonlinear dimensionality reduction by locally linear embedding.
\newblock {\em Science}, 290(5500):2323--2326, 2000.

\bibitem{bevilacqua2012low}
M.~Bevilacqua, A.~Roumy, C.~Guillemot, and M.-L.~A. Morel.
\newblock Low-complexity single-image super-resolution based on nonnegative
  neighbor embedding.
\newblock In {\em British Machine Vision Conference (BMVC)}, 2012.

\bibitem{chang2004super}
H.~Chang, D.-Y. Yeung, and Y.~Xiong.
\newblock Super-resolution through neighbor embedding.
\newblock In {\em IEEE Computer Society Conference on Computer Vision and
  Pattern Recognition}, volume~1, pages I--I. IEEE, 2004.

\bibitem{freeman2002example}
W.~T. Freeman, T.~R. Jones, and E.~C. Pasztor.
\newblock Example-based super-resolution.
\newblock {\em IEEE Comput. Graph. Appl.}, 22(2):56--65, 2002.

\bibitem{freeman2000learning}
W.~T. Freeman, E.~C. Pasztor, and O.~T. Carmichael.
\newblock Learning low-level vision.
\newblock {\em Int. J. Comput. Vis.}, 40(1):25--47, 2000.

\bibitem{lee2006efficient}
H.~Lee, A.~Battle, R.~Raina, and A.~Y. Ng.
\newblock Efficient sparse coding algorithms.
\newblock In {\em Proceedings of the 19th International Conference on Neural
  Information Processing Systems}, pages 801--808, 2006.

\bibitem{yang2008image}
J.~Yang, J.~Wright, T.~Huang, and Y.~Ma.
\newblock Image super-resolution as sparse representation of raw image patches.
\newblock In {\em IEEE conference on computer vision and pattern recognition},
  pages 1--8. IEEE, 2008.

\bibitem{lu2012geometry}
X.~Lu, H.~Yuan, P.~Yan, Y.~Yuan, and X.~Li.
\newblock Geometry constrained sparse coding for single image super-resolution.
\newblock In {\em IEEE Conference on Computer Vision and Pattern Recognition},
  pages 1648--1655. IEEE, 2012.

\bibitem{yang2010image}
J.~Yang, J.~Wright, T.~S. Huang, and Y.~Ma.
\newblock Image super-resolution via sparse representation.
\newblock {\em IEEE Trans. Image Process.}, 19(11):2861--2873, 2010.

\bibitem{zhang2012single}
K.~Zhang, X.~Gao, D.~Tao, and X.~Li.
\newblock Single image super-resolution with non-local means and steering
  kernel regression.
\newblock {\em IEEE Trans. Image Process.}, 21(11):4544--4556, 2012.

\bibitem{dong2014learning}
C.~Dong, C.~C. Loy, K.~He, and X.~Tang.
\newblock Learning a deep convolutional network for image super-resolution.
\newblock In {\em European conference on computer vision}, pages 184--199.
  Springer, 2014.

\bibitem{dong2016accelerating}
C.~Dong, C.~C. Loy, and X.~Tang.
\newblock Accelerating the super-resolution convolutional neural network.
\newblock In {\em European conference on computer vision}, pages 391--407.
  Springer, 2016.

\bibitem{yang2019deep}
W.~Yang, X.~Zhang, Y.~Tian, W.~Wang, J.-H. Xue, and Q.~Liao.
\newblock Deep learning for single image super-resolution: {A} brief review.
\newblock {\em IEEE Trans. Multimed.}, 21(12):3106--3121, 2019.

\bibitem{dong2015image}
C.~Dong, C.~C. Loy, K.~He, and X.~Tang.
\newblock Image super-resolution using deep convolutional networks.
\newblock {\em IEEE Trans. Pattern Anal. Mach. Intell.}, 38(2):295--307, 2015.

\bibitem{BNK2020}
S.~L. Brunton, B.~R. Noack, and P.~Koumoutsakos.
\newblock Machine learning for fluid mechanincs.
\newblock {\em Annu. Rev. Fluid Mech.}, 52:477--508, 2020.

\bibitem{BHT2020}
S.~L. Brunton, M.~S. Hemati, and K.~Taira.
\newblock Special issue on machine learning and data-driven methods in fluid
  dynamics.
\newblock {\em Theor. Comput. Fluid Dyn.}, 34:333--337, 2020.

\bibitem{BEF2019}
M.~P. Brenner, J.~D. Eldredge, and J.~B. Freund.
\newblock Perspective on machine learning for advancing fluid mechanics.
\newblock {\em Phys. Rev. Fluids}, 4:(100501), 2019.

\bibitem{DIX2019}
K.~Duraisamy, G.~Iaccarino, and H.~Xiao.
\newblock Turbulence modeling in the age of data.
\newblock {\em Annu. Rev. Fluid Mech.}, 51:357--377, 2019.

\bibitem{MSJC2019}
R.~Maulik, O.~San, J.~D. Jacob, and C.~Crick.
\newblock Sub-grid scale model classification and blending through deep
  learning.
\newblock {\em J. Fluid Mech.}, 870:784--812, 2019.

\bibitem{LKT2016}
J.~Ling, A.~Kurzawski, and J~Templeton.
\newblock Reynolds averaged turbulence modelling using deep neural networks
  with embedded invariance.
\newblock {\em J. Fluid Mech.}, 807:155--166, 2016.

\bibitem{novati2021automating}
G.~Novati, H.~L. de~Laroussilhe, and P.~Koumoutsakos.
\newblock Automating turbulence modelling by multi-agent reinforcement
  learning.
\newblock {\em Nat. Mach. Intell.}, 3(1):87--96, 2021.

\bibitem{bae2022scientific}
H.~J. Bae and P.~Koumoutsakos.
\newblock Scientific multi-agent reinforcement learning for wall-models of
  turbulent flows.
\newblock {\em Nat. Commun.}, 13(1):1--9, 2022.

\bibitem{LY2019}
S.~Lee and D.~You.
\newblock Data-driven prediction of unsteady flow fields over a circular
  cylinder using deep learning.
\newblock {\em J. Fluid Mech.}, 879:217--254, 2019.

\bibitem{callaham2022empirical}
J.~L. Callaham, G.~Rigas, J.-C. Loiseau, and S.~L. Brunton.
\newblock An empirical mean-field model of symmetry-breaking in a turbulent
  wake.
\newblock {\em Sci. Adv.}, 8(19):eabm4786, 2022.

\bibitem{san2018neural}
O.~San and R.~Maulik.
\newblock Neural network closures for nonlinear model order reduction.
\newblock {\em Adv. Comput. Math.}, 44(6):1717--1750, 2018.

\bibitem{fukami2020sparse}
K.~Fukami, T.~Murata, K.~Zhang, and K.~Fukagata.
\newblock Sparse identification of nonlinear dynamics with low-dimensionalized
  flow representations.
\newblock {\em J. Fluid Mech.}, 926:A10, 2021.

\bibitem{SGASV2019}
P.~A. Srinivasan, L.~Guastoni, H.~Azizpour, P.~Schlatter, and R.~Vinuesa.
\newblock Predictions of turbulent shear flows using deep neural networks.
\newblock {\em Phys. Rev. Fluids}, 4:054603, 2019.

\bibitem{RPCA2020}
I.~Scherl, B.~Strom, J.~K. Shang, O.~Williams, B.~L. Polagye, and S.~L.
  Brunton.
\newblock Robust principal component analysis for modal decomposition of
  corrupt fluid flows.
\newblock {\em Phys. Rev. Fluids}, 5:054401, 2020.

\bibitem{manohar2018data}
K.~Manohar, B.~W. Brunton, J.~N. Kutz, and S.~L. Brunton.
\newblock Data-driven sparse sensor placement for reconstruction: Demonstrating
  the benefits of exploiting known patterns.
\newblock {\em IEEE Control Systems Magazine}, 38(3):63--86, 2018.

\bibitem{FFT2020}
K.~Fukami, K.~Fukagata, and K.~Taira.
\newblock Assessment of supervised machine learning for fluid flows.
\newblock {\em Theor. Comput. Fluid Dyn.}, 34(4):497--519, 2020.

\bibitem{KKL2023}
H.~Kim, J.~Kim, and C.~Lee.
\newblock Interpretable deep learning for prediction of prandtl number effect
  in turbulent heat transfer.
\newblock {\em J. Fluid Mech.}, 955:A14, 2023.

\bibitem{rabault2019artificial}
J.~Rabault, M.~Kuchta, A.~Jensen, U.~R{\'e}glade, and N.~Cerardi.
\newblock Artificial neural networks trained through deep reinforcement
  learning discover control strategies for active flow control.
\newblock {\em J. Fluid Mech,}, 865:281--302, 2019.

\bibitem{bieker2020deep}
K.~Bieker, S.~Peitz, S.~L. Brunton, J.~N. Kutz, and M.~Dellnitz.
\newblock Deep model predictive flow control with limited sensor data and
  online learning.
\newblock {\em Theor. Comput. Fluid Dyn.}, 34(4):577--591, 2020.

\bibitem{zhou2020artificial}
Y.~Zhou, D.~Fan, B.~Zhang, R.~Li, and B.~R. Noack.
\newblock Artificial intelligence control of a turbulent jet.
\newblock {\em J. Fluid Mech.}, 897:A27, 2020.

\bibitem{paris2021robust}
R.~Paris, S.~Beneddine, and J.~Dandois.
\newblock Robust flow control and optimal sensor placement using deep
  reinforcement learning.
\newblock {\em J. Fluid Mech.}, 913:A25, 2021.

\bibitem{park2020machine}
J.~Park and H.~Choi.
\newblock Machine-learning-based feedback control for drag reduction in a
  turbulent channel flow.
\newblock {\em J. Fluid Mech.}, 904:A24, 2020.

\bibitem{ghraieb2021single}
H.~Ghraieb, J.~Viquerat, A.~Larcher, P.~Meliga, and E.~Hachem.
\newblock Single-step deep reinforcement learning for open-loop control of
  laminar and turbulent flows.
\newblock {\em Phys. Rev. Fluids}, 6(5):053902, 2021.

\bibitem{fukami2021global}
K.~Fukami, R.~Maulik, N.~Ramachandra, K.~Fukagata, and K.~Taira.
\newblock Global field reconstruction from sparse sensors with {V}oronoi
  tessellation-assisted deep learning.
\newblock {\em Nat. Mach. Intell.}, 3(11):945--951, 2021.

\bibitem{guemes2022super}
A.~G{\"u}emes, C.~S. Vila, and S.~Discetti.
\newblock Super-resolution generative adversarial networks of randomly-seeded
  fields.
\newblock {\em Nat. Mach. Intell.}, 4:1165--1173, 2022.

\bibitem{sun2020physics}
L.~Sun and J.-X. Wang.
\newblock Physics-constrained {B}ayesian neural network for fluid flow
  reconstruction with sparse and noisy data.
\newblock {\em Theor. Appl. Mech. Lett.}, 10(3):161--169, 2020.

\bibitem{gao2021super}
H.~Gao, L.~Sun, and J.-X. Wang.
\newblock Super-resolution and denoising of fluid flow using physics-informed
  convolutional neural networks without high-resolution labels.
\newblock {\em Phys. Fluids}, 33(7):073603, 2021.

\bibitem{fathi2020super}
M.~F. Fathi, I.~Perez-Raya, A.~Baghaie, P.~Berg, G.~Janiga, A.~Arzani, and
  R.~M. D’Souza.
\newblock Super-resolution and denoising of {4D}-flow {MRI} using
  physics-informed deep neural nets.
\newblock {\em Comput. Methods Programs Biomed.}, 197:105729, 2020.

\bibitem{vlasenko2009superresolution}
A.~Vlasenko and C.~Schn{\"o}rr.
\newblock Superresolution and denoising of 3{D} fluid flow estimates.
\newblock In {\em Joint Pattern Recognition Symposium}, pages 482--491.
  Springer, 2009.

\bibitem{pradhan2021variational}
A.~Pradhan and K.~Duraisamy.
\newblock Variational multi-scale super-resolution: A data-driven approach for
  reconstruction and predictive modeling of unresolved physics.
\newblock {\em arXiv:2101.09839}, 2021.

\bibitem{bode2021using}
M.~Bode, M.~Gauding, Z.~Lian, D.~Denker, M.~Davidovic, K.~Kleinheinz,
  J.~Jitsev, and H.~Pitsch.
\newblock Using physics-informed enhanced super-resolution generative
  adversarial networks for subfilter modeling in turbulent reactive flows.
\newblock {\em Proc. Combust. Inst.}, 38(2):2617--2625, 2021.

\bibitem{xie2018tempogan}
Y.~Xie, E.~Franz, M.~Chu, and N.~Thuerey.
\newblock tempogan: {A} temporally coherent, volumetric {GAN} for
  super-resolution fluid flow.
\newblock {\em ACM Trans. Graph.}, 37(4):1--15, 2018.

\bibitem{FFT2019a}
K.~Fukami, K.~Fukagata, and K.~Taira.
\newblock Super-resolution reconstruction of turbulent flows with machine
  learning.
\newblock {\em J. Fluid Mech.}, 870:106--120, 2019.

\bibitem{erichson2020shallow}
N.~B. Erichson, L.~Mathelin, Z.~Yao, S.~L. Brunton, M.~W. Mahoney, and J.~N.
  Kutz.
\newblock Shallow neural networks for fluid flow reconstruction with limited
  sensors.
\newblock {\em Proc. Roy. Soc. A}, 476(2238):20200097, 2020.

\bibitem{bode2019deep}
M.~Bode, M.~Gauding, K.~Kleinheinz, and H.~Pitsch.
\newblock Deep learning at scale for subgrid modeling in turbulent flows:
  regression and reconstruction.
\newblock In {\em International Conference on High Performance Computing},
  pages 541--560. Springer, 2019.

\bibitem{obiols2021surfnet}
O.~Obiols-Sales, A.~Vishnu, N.~P. Malaya, and A.~Chandramowlishwaran.
\newblock {SURFNet}: {S}uper-resolution of turbulent flows with transfer
  learning using small datasets.
\newblock In {\em IEEE 30th International Conference on Parallel Architectures
  and Compilation Techniques (PACT)}, pages 331--344. IEEE, 2021.

\bibitem{liu2020deep}
B.~Liu, J.~Tang, H.~Huang, and X.-Y. Lu.
\newblock Deep learning methods for super-resolution reconstruction of
  turbulent flows.
\newblock {\em Phys. Fluids}, 32(2):025105, 2020.

\bibitem{kim2021unsupervised}
H.~Kim, J.~Kim, S.~Won, and C.~Lee.
\newblock Unsupervised deep learning for super-resolution reconstruction of
  turbulence.
\newblock {\em J. Fluid Mech.}, 910:A29, 2021.

\bibitem{zhou2022neural}
X.-H. Zhou, J.~E. McClure, C.~Chen, and H.~Xiao.
\newblock Neural network--based pore flow field prediction in porous media
  using super resolution.
\newblock {\em Phys. Rev. Fluids}, 7(7):074302, 2022.

\bibitem{guemes2021coarse}
A.~G{\"u}emes, S.~Discetti, A.~Ianiro, B.~Sirmacek, H.~Azizpour, and
  R.~Vinuesa.
\newblock From coarse wall measurements to turbulent velocity fields through
  deep learning.
\newblock {\em Phys. Fluids}, 33(7):075121, 2021.

\bibitem{yousif2021high}
M.~Z. Yousif, L.~Yu, and H.-C. Lim.
\newblock High-fidelity reconstruction of turbulent flow from spatially limited
  data using enhanced super-resolution generative adversarial network.
\newblock {\em Phys. Fluids}, 33(12):125119, 2021.

\bibitem{NG2020}
N.~J. Nair and A.~Goza.
\newblock Leveraging reduced-order models for state estimation using deep
  learning.
\newblock {\em J. Fluid Mech.}, 897:R1, 2020.

\bibitem{RHW1986}
D.~E. Rumelhart, G.~E. Hinton, and R.~J. Williams.
\newblock Learning representations by back-propagation errors.
\newblock {\em Nature}, 322:533---536, 1986.

\bibitem{Kingma2014}
D.~P. Kingma and J.~Ba.
\newblock Adam: A method for stochastic optimization.
\newblock {\em {arXiv:1412.6980}\hspace{-0.3em}}, 2014.

\bibitem{williams2022data}
J.~Williams, O.~Zahn, and J.~N. Kutz.
\newblock Data-driven sensor placement with shallow decoder networks.
\newblock {\em arXiv:2202.05330}, 2022.

\bibitem{domingos2012few}
P.~Domingos.
\newblock A few useful things to know about machine learning.
\newblock {\em Commun. ACM}, 55(10):78--87, 2012.

\bibitem{LBBH1998}
Y.~LeCun, L.~Bottou, Y.~Bengio, and P.~Haffner.
\newblock Gradient-based learning applied to document recognition.
\newblock {\em Proc. IEEE}, 86(11):2278--2324, 1998.

\bibitem{morimoto2021convolutional}
M.~Morimoto, K.~Fukami, K.~Zhang, A.~G. Nair, and K.~Fukagata.
\newblock Convolutional neural networks for fluid flow analysis: toward
  effective metamodeling and low dimensionalization.
\newblock {\em Theor. Comput. Fluid Dyn.}, 35(5):633--658, 2021.

\bibitem{NF2022}
T.~Nakamura and K.~Fukagata.
\newblock Robust training approach of neural networks for fluid flow state
  estimations.
\newblock {\em Int. J. Heat Fluid Flow}, 96:108997, 2022.

\bibitem{wurster2022deep}
S.~W. Wurster, H.~Guo, H.-W. Shen, T.~Peterka, and J.~Xu.
\newblock Deep hierarchical super resolution for scientific data.
\newblock {\em IEEE Trans. Vis. Comput. Graph.}, 2022.

\bibitem{romano2016raisr}
Y.~Romano, J.~Isidoro, and P.~Milanfar.
\newblock {RAISR}: rapid and accurate image super resolution.
\newblock {\em IEEE Trans. Comput. Imaging}, 3(1):110--125, 2016.

\bibitem{goodfellow2020generative}
I.~Goodfellow, J.~Pouget-Abadie, M.~Mirza, B.~Xu, D.~Warde-Farley, S.~Ozair,
  A.~Courville, and Y.~Bengio.
\newblock Generative adversarial networks.
\newblock {\em Commun. ACM}, 63(11):139--144, 2020.

\bibitem{MTK2023}
S.~Maejima, K.~Tanino, and S.~Kawai.
\newblock Unsupervised machine-learning-based sub-grid scale modeling for
  coarse-grid {LES}.
\newblock {\em \rm In review}, 2023.

\bibitem{rumelhart1985feature}
D.~E. Rumelhart and D.~Zipser.
\newblock Feature discovery by competitive learning.
\newblock {\em Cogn. Sci.}, 9(1):75--112, 1985.

\bibitem{lagaris1998artificial}
I.~E. Lagaris, A.~Likas, and D.~I. Fotiadis.
\newblock Artificial neural networks for solving ordinary and partial
  differential equations.
\newblock {\em IEEE Trans. Neural Netw.}, 9(5):987--1000, 1998.

\bibitem{raissi2019physics}
M.~Raissi, P.~Perdikaris, and G.~E. Karniadakis.
\newblock Physics-informed neural networks: A deep learning framework for
  solving forward and inverse problems involving nonlinear partial differential
  equations.
\newblock {\em J. Comput. Phys.}, 378:686--707, 2019.

\bibitem{karniadakis2021physics}
G.~E. Karniadakis, I.~G. Kevrekidis, L.~Lu, P.~Perdikaris, S.~Wang, and
  L.~Yang.
\newblock Physics-informed machine learning.
\newblock {\em Nat. Rev. Phys.}, 3(6):422--440, 2021.

\bibitem{cai2022physics}
S.~Cai, Z.~Mao, Z.~Wang, M.~Yin, and G.~E. Karniadakis.
\newblock Physics-informed neural networks {(PINNs)} for fluid mechanics: {A}
  review.
\newblock {\em Acta Mech. Sin.}, pages 1--12, 2022.

\bibitem{zhu2009introduction}
X.~Zhu and A.~B. Goldberg.
\newblock {\em Introduction to semi-supervised learning}, volume~3.
\newblock Morgan \& Claypool Publishers, 2009.

\bibitem{FFT2019tsfp}
K.~Fukami, K.~Fukagata, and K.~Taira.
\newblock Super-resolution analysis with machine learning for low-resolution
  flow data.
\newblock In {\em 11th International Symposium on Turbulence and Shear Flow
  Phenomena (TSFP11), Southampton, UK}, number 208, 2019.

\bibitem{FFT2021b}
K.~Fukami, K.~Fukagata, and K.~Taira.
\newblock Machine-learning-based spatio-temporal super resolution
  reconstruction of turbulent flows.
\newblock {\em J. Fluid Mech.}, 909(A9), 2021.

\bibitem{he2016deep}
K.~He, X.~Zhang, S.~Ren, and J.~Sun.
\newblock Deep residual learning for image recognition.
\newblock In {\em Proceedings of the IEEE conference on computer vision and
  pattern recognition}, pages 770--778, 2016.

\bibitem{pan2009survey}
S.~J. Pan and Q.~Yang.
\newblock A survey on transfer learning.
\newblock {\em IEEE Trans. Knowl. Data Eng.}, 22(10):1345--1359, 2009.

\bibitem{guastoni2021convolutional}
L.~Guastoni, A.~G{\"u}emes, A.~Ianiro, S.~Discetti, P.~Schlatter, H.~Azizpour,
  and R.~Vinuesa.
\newblock Convolutional-network models to predict wall-bounded turbulence from
  wall quantities.
\newblock {\em J. Fluid Mech.}, 928:A27, 2021.

\bibitem{LPBK2020}
Y.~Liu, C.~Ponce, S.~L. Brunton, and J.~N. Kutz.
\newblock Multiresolution convolutional autoencoders.
\newblock {\em J. Comput. Phys.}, page 111801, 2022.

\bibitem{pant2020deep}
P.~Pant and A.~B. Farimani.
\newblock Deep learning for efficient reconstruction of high-resolution
  turbulent {DNS} data.
\newblock {\em arXiv:2010.11348}, 2020.

\bibitem{kong2020deep}
C.~Kong, J.-T. Chang, Y.-F. Li, and R.-Y. Chen.
\newblock Deep learning methods for super-resolution reconstruction of
  temperature fields in a supersonic combustor.
\newblock {\em AIP Adv.}, 10(11):115021, 2020.

\bibitem{matsuo2021supervised}
M.~Matsuo, T.~Nakamura, M.~Morimoto, K.~Fukami, and K.~Fukagata.
\newblock Supervised convolutional network for three-dimensional fluid data
  reconstruction from sectional flow fields with adaptive super-resolution
  assistance.
\newblock {\em arXiv:2103.09020}, 2021.

\bibitem{LRT2019}
Y.~Li, D.~Roblek, and M.~Tagliasacchi.
\newblock From here to there: {V}ideo inbetweening using 3{D} convolutions.
\newblock {\em {arXiv:1905.10240}\hspace{-0.3em}}, 2019.

\bibitem{AS2022}
A.~Shrivastava R.~Arora.
\newblock Spatio-temporal super-resolution of dynamical systems using
  physics-informed deep-learning.
\newblock {\em AAAI 2023: Workshop on AI to Accelerate Science and Engineering
  (AI2ASE)}, 2022.

\bibitem{KL2020}
J.~Kim and C.~Lee.
\newblock Prediction of turbulent heat transfer using convolutional neural
  networks.
\newblock {\em J. Fluid Mech.}, 882:A18, 2020.

\bibitem{morimoto2022generalization}
M.~Morimoto, K.~Fukami, K.~Zhang, and K.~Fukagata.
\newblock Generalization techniques of neural networks for fluid flow
  estimation.
\newblock {\em Neural Comput. App.}, 34(5):3647--3669, 2022.

\bibitem{onishi2019super}
R.~Onishi, D.~Sugiyama, and K.~Matsuda.
\newblock Super-resolution simulation for real-time prediction of urban
  micrometeorology.
\newblock {\em {SOLA}}, 15:178--182, 2019.

\bibitem{yasuda2022super}
Y.~Yasuda, R.~Onishi, Y.~Hirokawa, D.~Kolomenskiy, and D.~Sugiyama.
\newblock Super-resolution of near-surface temperature utilizing physical
  quantities for real-time prediction of urban micrometeorology.
\newblock {\em Build. Environ.}, 209:108597, 2022.

\bibitem{hu2018squeeze}
J.~Hu, L.~Shen, and G.~Sun.
\newblock Squeeze-and-excitation networks.
\newblock In {\em Proceedings of the IEEE conference on computer vision and
  pattern recognition}, pages 7132--7141, 2018.

\bibitem{MLPIV2023}
S.~Discetti and Y.~Liu.
\newblock Machine learning for flow field measurements: a perspective.
\newblock {\em Meas. Sci. Technol.}, 34:021001, 2023.

\bibitem{DHLK2019}
Z.~Deng, C.~He, Y.~Liu, and K.~C. Kim.
\newblock Super-resolution reconstruction of turbulent velocity fields using a
  generative adversarial network-based artificial intelligence framework.
\newblock {\em Phys. Fluids}, 31:125111, 2019.

\bibitem{wang2020predicting}
H.~Wang, Z.~Yang, B.~Li, and S.~Wang.
\newblock Predicting the near-wall velocity of wall turbulence using a neural
  network for particle image velocimetry.
\newblock {\em Phys. Fluids}, 32(11):115105, 2020.

\bibitem{MFF2019}
T.~Murata, K.~Fukami, and K.~Fukagata.
\newblock Nonlinear mode decomposition with convolutional neural networks for
  fluid dynamics.
\newblock {\em J. Fluid Mech.}, 882:A13, 2020.

\bibitem{adrian2005twenty}
R.~J. Adrian.
\newblock Twenty years of particle image velocimetry.
\newblock {\em Exp. Fluids}, 39(2):159--169, 2005.

\bibitem{CZXG2019}
S.~Cai, S.~Zhou, C.~Xu, and Q.~Gao.
\newblock Dense motion estimation of particle images via a convolutional neural
  network.
\newblock {\em Exp. Fluids}, 60:60--73, 2019.

\bibitem{deconv2015}
A.~Dosovitskiy, P.~Fischer, J.~T. Springenberg, M.~Riedmiller, and T.~Brox.
\newblock Discriminative unsupervised feature learning with exemplar
  convolutional neural networks.
\newblock {\em IEEE Trans. Pattern Anal. Mach. Intell.}, 38, 2019.

\bibitem{majewski2020developing}
W.~Majewski, R.~Wei, and V.~Kumar.
\newblock Developing particle image velocimetry software based on a deep neural
  network.
\newblock {\em J. Flow Vis. Image Process.}, 27(4), 2020.

\bibitem{morimoto2021experimental}
M.~Morimoto, K.~Fukami, and K.~Fukagata.
\newblock Experimental velocity data estimation for imperfect particle images
  using machine learning.
\newblock {\em Phys. Fluids}, 33(8):087121, 2021.

\bibitem{FT2022b}
K.~Fukami and K.~Taira.
\newblock Learning the nonlinear manifold of extreme aerodynamics.
\newblock {\em NeurIPS2022}, 2022.

\bibitem{FNF2020}
K.~Fukami, T.~Nakamura, and K.~Fukagata.
\newblock Convolutional neural network based hierarchical autoencoder for
  nonlinear mode decomposition of fluid field data.
\newblock {\em Phys. Fluids}, 32:095110, 2020.

\bibitem{ricardo_VAE2021}
H.~Eivazi, S.~Le~Clainche, S.~Hoyas, and R.~Vinuesa.
\newblock Towards extraction of orthogonal and parsimonious non-linear modes
  from turbulent flows.
\newblock {\em Expert Syst. Appl.}, 202:117038, 2022.

\bibitem{FHNMF2020}
K.~Fukami, K.~Hasegawa, T.~Nakamura, M.~Morimoto, and K.~Fukagata.
\newblock Model order reduction with neural networks: Application to laminar
  and turbulent flows.
\newblock {\em SN Comput. Sci.}, 2:467, 2021.

\bibitem{linot2020deep}
A.~J. Linot and M.~D. Graham.
\newblock Deep learning to discover and predict dynamics on an inertial
  manifold.
\newblock {\em Phys. Rev. E}, 101(6):062209, 2020.

\bibitem{FT2023}
K.~Fukami and K.~Taira.
\newblock Grasping extreme aerodynamics on a low-dimensional manifold.
\newblock {\em {\rm arXiv:2305.08024}}, 2023.

\bibitem{wu2020comprehensive}
Z.~Wu, S.~Pan, F.~Chen, G.~Long, C.~Zhang, and S.~Y. Philip.
\newblock A comprehensive survey on graph neural networks.
\newblock {\em IEEE Trans. Neural Netw. Learn. Syst.}, 32(1):4--24, 2020.

\bibitem{carter2021data}
D.~W. Carter, F.~De~Voogt, R.~Soares, and B.~Ganapathisubramani.
\newblock Data-driven sparse reconstruction of flow over a stalled aerofoil
  using experimental data.
\newblock {\em Data-Centric Eng.}, 2, 2021.

\bibitem{giannopoulos2020data}
A.~Giannopoulos and J.-L. Aider.
\newblock Data-driven order reduction and velocity field reconstruction using
  neural networks: The case of a turbulent boundary layer.
\newblock {\em Phys. Fluids}, 32(9):095117, 2020.

\bibitem{MFRFT2020}
R.~Maulik, K.~Fukami, N.~Ramachandra, K.~Fukagata, and K.~Taira.
\newblock Probabilistic neural networks for fluid flow surrogate modeling and
  data recovery.
\newblock {\em Phys. Rev. Fluids}, 5:104401, 2020.

\bibitem{ES1995}
R.~Everson and L.~Sirovich.
\newblock Karhunen--{L}oeve procedure for gappy data.
\newblock {\em J. Opt. Soc. Am.}, 12(8):1657--1664, 1995.

\bibitem{BDW2004}
T.~Bui-Thanh, M.~Damodaran, and K.~Willcox.
\newblock Aerodynamic data reconstruction and inverse design using proper
  orthogonal decomposition.
\newblock {\em AIAA J.}, 42(8):1505--1516, 2004.

\bibitem{adrian1988stochastic}
R.~J. Adrian and P.~Moin.
\newblock Stochastic estimation of organized turbulent structure: homogeneous
  shear flow.
\newblock {\em J. Fluid Mech.}, 190:531--559, 1988.

\bibitem{manohar2022sparse}
K.~H. Manohar, C.~Morton, and P.~Ziad{\'e}.
\newblock Sparse sensor-based cylinder flow estimation using artificial neural
  networks.
\newblock {\em Phys. Rev. Fluids}, 7(2):024707, 2022.

\bibitem{HS1997}
S.~Hochreiter and J.~Schmidhuber.
\newblock Long short-term memory.
\newblock {\em Neural Comput.}, 9:1735--1780, 1997.

\bibitem{dubois2022machine}
P.~Dubois, T.~Gomez, L.~Planckaert, and L.~Perret.
\newblock Machine learning for fluid flow reconstruction from limited
  measurements.
\newblock {\em J. Comput. Phys.}, 448:110733, 2022.

\bibitem{Lumely1967}
J.~L. Lumley.
\newblock The structure of inhomogeneous turbulent flows.
\newblock In A.~M. Yaglom and V.~I. Tatarski, editors, {\em Atmospheric
  turbulence and radio wave propagation}. Nauka, 1967.

\bibitem{Holmes}
P.~Holmes, J.L. Lumley, G.~Berkooz, and C.~W. Rowley.
\newblock {\em Turbulence, Coherent Structures, Dynamical Systems and
  Symmetry}.
\newblock Cambridge Univ. Press, 2nd edition, 2012.

\bibitem{TBDRCMSGTU2017}
K.~Taira, S.~L. Brunton, S.~T.~M. Dawson, C.~W. Rowley, T.~Colonius, B.~J.
  McKeon, O.~T. Schmidt, S.~Gordeyev, V.~Theofilis, and L.~S. Ukeiley.
\newblock Modal analysis of fluid flows: An overview.
\newblock {\em AIAA J.}, 55(12):4013--4041, 2017.

\bibitem{HS2006}
G.~E. Hinton and R.~R. Salakhutdinov.
\newblock Reducing the dimensionality of data with neural networks.
\newblock {\em Science}, 313(5786):504--507, 2006.

\bibitem{rezende2014stochastic}
D.~J. Rezende, S.~Mohamed, and D.~Wierstra.
\newblock Stochastic backpropagation and approximate inference in deep
  generative models.
\newblock In {\em International conference on machine learning}, pages
  1278--1286. PMLR, 2014.

\bibitem{smola2004tutorial}
A.~J. Smola and B.~Sch{\"o}lkopf.
\newblock A tutorial on support vector regression.
\newblock {\em Stat. Comput.}, 14(3):199--222, 2004.

\bibitem{friedman2001greedy}
J.~H. Friedman.
\newblock Greedy function approximation: a gradient boosting machine.
\newblock {\em Ann. Stat.}, pages 1189--1232, 2001.

\bibitem{BPK2016a}
S.~L. Brunton, J.~L. Proctor, and J.~N. Kutz.
\newblock Discovering governing equations from data by sparse identification of
  nonlinear dynamical systems.
\newblock {\em Proc. Natl. Acad. Sci. U.S.A.}, 113(15):3932--3937, 2016.

\bibitem{callaham2019robust}
J.~L. Callaham, K.~Maeda, and S.~L. Brunton.
\newblock Robust flow reconstruction from limited measurements via sparse
  representation.
\newblock {\em Phys. Rev. Fluids}, 4(10):103907, 2019.

\bibitem{MFMVF2022}
M.~Morimoto, K.~Fukami, R.~Maulik, R.~Vinuesa, and K.~Fukagata.
\newblock Assessments of epistemic uncertainty using gaussian stochastic weight
  averaging for fluid-flow regression.
\newblock {\em Phys. D: Nonlinear Phenom.}, 440:133454, 2022.

\bibitem{ZFAT2022}
Y.~Zhong, K.~Fukami, B.~An, and K.~Taira.
\newblock Machine-learning-based reconstruction of transient vortex-airfoil
  wake interaction.
\newblock {\em \rm AIAA paper, 2022-3244}, 2022.

\bibitem{ZFAT2023}
Y.~Zhong, K.~Fukami, B.~An, and K.~Taira.
\newblock Sparse sensor reconstruction of vortex-impinged airfoil wake with
  machine learning.
\newblock {\em Theor. Comput. Fluid Dyn.}, 2023.

\bibitem{lee2022predicting}
S.~Lee, J.~Yang, P.~Forooghi, A.~Stroh, and S.~Bagheri.
\newblock Predicting drag on rough surfaces by transfer learning of empirical
  correlations.
\newblock {\em J. Fluid Mech.}, 933:A18, 2022.

\bibitem{raissi2020hidden}
M.~Raissi, A.~Yazdani, and G.~E. Karniadakis.
\newblock Hidden fluid mechanics: Learning velocity and pressure fields from
  flow visualizations.
\newblock {\em Science}, 367(6481):1026--1030, 2020.

\bibitem{yousif2022deep}
M.~Z. Yousif, L.~Yu, S.~Hoyas, R.~Vinuesa, and H.~C. Lim.
\newblock A deep-learning approach for reconstructing {3D} turbulent flows from
  {2D} observation data.
\newblock {\em arXiv:2208.05754}, 2022.

\bibitem{wang2018esrgan}
X.~Wang, K.~Yu, S.~Wu, J.~Gu, Y.~Liu, C.~Dong, Y.~Qiao, and C.~Change~L.
\newblock {ESRGAN}: {E}nhanced super-resolution generative adversarial
  networks.
\newblock In {\em Proceedings of the European conference on computer vision
  (ECCV) workshops}, pages 63--79, 2018.

\bibitem{HFMF2020a}
K.~Hasegawa, K.~Fukami, T.~Murata, and K.~Fukagata.
\newblock Machine-learning-based reduced-order modeling for unsteady flows
  around bluff bodies of various shapes.
\newblock {\em Theor. Comput. Fluid Dyn.}, 34(4):367--388, 2020.

\bibitem{HFMF2020b}
K.~Hasegawa, K.~Fukami, T.~Murata, and K.~Fukagata.
\newblock {CNN-LSTM} based reduced order modeling of two-dimensional unsteady
  flows around a circular cylinder at different {Reynolds} numbers.
\newblock {\em Fluid Dyn. Res.}, 52(6):065501, 2020.

\bibitem{esmaeilzadeh2020meshfreeflownet}
S.~Esmaeilzadeh, K.~Azizzadenesheli, K.~Kashinath, M.~Mustafa, H.~A. Tchelepi,
  P.~Marcus, M.~Prabhat, Anandkumar., et~al.
\newblock {MeshfreeFlowNet: A} physics-constrained deep continuous space-time
  super-resolution framework.
\newblock In {\em SC20: International Conference for High Performance
  Computing, Networking, Storage and Analysis}, pages 1--15. IEEE, 2020.

\bibitem{TransFlowNet2022}
X.~Wang, S.~Zhu, Y.~Guo, P.~Han, Y.~Wang, Z.~Wei, and X.~Jin.
\newblock {TransFlowNet: A} physics-constrained transformer framework for
  spatio-temporal super-resolution of flow simulations.
\newblock {\em J. Comput. Sci.}, 65(101906), 2022.

\bibitem{zhu2017unpaired}
J.-Y. Zhu, T.~Park, P.~Isola, and A.~A. Efros.
\newblock Unpaired image-to-image translation using cycle-consistent
  adversarial networks.
\newblock In {\em Proceedings of the IEEE international conference on computer
  vision}, pages 2223--2232, 2017.

\bibitem{psaros2022meta}
A.~F. Psaros, K.~Kawaguchi, and G.~E. Karniadakis.
\newblock Meta-learning {PINN} loss functions.
\newblock {\em J. Comput. Phys.}, 458:111121, 2022.

\bibitem{zhang2022towards}
B.~Zhang, R.~Ooka, H.~Kikumoto, C.~Hu, and K.~T. Tim.
\newblock Towards real-time prediction of velocity field around a building
  using generative adversarial networks based on the surface pressure from
  sparse sensor networks.
\newblock {\em J. Wind. Eng. Ind.}, 231:105243, 2022.

\bibitem{ioffe2015batch}
S.~Ioffe and C.~Szegedy.
\newblock Batch normalization: {A}ccelerating deep network training by reducing
  internal covariate shift.
\newblock In {\em International conference on machine learning}, pages
  448--456, 2015.

\bibitem{wurster2021deep}
S.~W. Wurster, H.-W. Shen, H.~Guo, T.~Peterka, M.~Raj, and J.~Xu.
\newblock Deep hierarchical super-resolution for scientific data reduction and
  visualization.
\newblock {\em arXiv:2107.00462}, 2021.

\bibitem{BMT2001}
T.~R. Bewley, P.~Moin, and R.~Temam.
\newblock {DNS}-based predictive control of turbulence: an optimal benchmark
  for feedback algorithms.
\newblock {\em J. Fluid Mech.}, 447:179--225, 2001.

\bibitem{CHBH2006}
M.~Chevalier, J.~H{\oe}pffner, T.~R. Bewley, and D.~S. Henningson.
\newblock State estimation in wall-bounded flow systems. {P}art 2. {T}urbulent
  flows.
\newblock {\em J. Fluid Mech.}, 552:167--187, 2006.

\bibitem{CCB2011}
C.~H. Colburn, J.~B. Cessna, and T.~R. Bewley.
\newblock State estimation in wall-bounded flow systems. {P}art 3. {T}he
  ensemble kalman filter.
\newblock {\em J. Fluid Mech.}, 682:289--303, 2011.

\bibitem{SH2017}
T.~Suzuki and Y.~Hasegawa.
\newblock Estimation of turbulent channel flow at {${Re}_{\tau} = 100$} based
  on the wall measurement using a simple sequential approach.
\newblock {\em J. Fluid Mech.}, 830:760--796, 2006.

\bibitem{yousif2022super}
M.~Z. Yousif, L.~Yu, and H.-C. Lim.
\newblock Super-resolution reconstruction of turbulent flow fields at various
  {R}eynolds numbers based on generative adversarial networks.
\newblock {\em Phys. Fluids}, 34(1):015130, 2022.

\bibitem{xu2020data}
W.~Xu, W.~Luo, Y.~Wang, and Y.~You.
\newblock Data-driven three-dimensional super-resolution imaging of a turbulent
  jet flame using a generative adversarial network.
\newblock {\em Appl. Opt.}, 59(19):5729--5736, 2020.

\bibitem{hassanaly2022adversarial}
M.~Hassanaly, A.~Glaws, K.~Stengel, and R.~N. King.
\newblock Adversarial sampling of unknown and high-dimensional conditional
  distributions.
\newblock {\em J. Comput. Phys.}, 450:110853, 2022.

\bibitem{ledig2017photo}
C.~Ledig, L.~Theis, F.~Husz{\'a}r, J.~Caballero, A.~Cunningham, A.~Acosta,
  A.~Aitken, A.~Tejani, J.~Totz, Z.~Wang, and W.~Shi.
\newblock Photo-realistic single image super-resolution using a generative
  adversarial network.
\newblock In {\em Proceedings of the IEEE conference on computer vision and
  pattern recognition}, pages 4681--4690, 2017.

\bibitem{stengel2020adversarial}
K.~Stengel, A.~Glaws, D.~Hettinger, and R.~N. King.
\newblock Adversarial super-resolution of climatological wind and solar data.
\newblock {\em Proc. Natl. Acad. Sci. U.S.A.}, 117(29):16805--16815, 2020.

\bibitem{yang2019diversity}
D.~Yang, S.~Hong, Y.~Jang, T.~Zhao, and H.~Lee.
\newblock Diversity-sensitive conditional generative adversarial networks.
\newblock In {\em 7th International Conference on Learning Representations,
  ICLR 2019}. International Conference on Learning Representations, ICLR, 2019.

\bibitem{TNB2016}
K.~Taira, A.~G. Nair, and S.~L. Brunton.
\newblock Network structure of two-dimensional decaying isotropic turbulence.
\newblock {\em J. Fluid Mech.}, 795:R2, 2016.

\bibitem{DQHG2018}
X.~Du, X.~Qu, Y.~He, and D.~Guo.
\newblock Single image super-resolution based on multi-scale competitive
  convolutional neural network.
\newblock {\em Sensors}, 789(18):1--17, 2018.

\bibitem{NH2010}
V.~Nair and G.~E. Hinton.
\newblock Rectified linear units improve restricted boltzmann machines.
\newblock {\em In Proc. 27th International Conference on Machine Learning},
  2010.

\bibitem{ren2022physics}
P.~Ren, C.~Rao, Y.~Liu, Z.~Ma, Q.~Wang, J.-X. Wang, and H.~Sun.
\newblock Physics-informed deep super-resolution for spatiotemporal data.
\newblock {\em arXiv:2208.01462}, 2022.

\bibitem{kashefi2021point}
A.~Kashefi, D.~Rempe, and L.~J. Guibas.
\newblock A point-cloud deep learning framework for prediction of fluid flow
  fields on irregular geometries.
\newblock {\em Phys. Fluids}, 33(2):027104, 2021.

\bibitem{qi2017pointnet}
C.~R. Qi, H.~Su, K.~Mo, and L.~J. Guibas.
\newblock {PointNet}: Deep learning on point sets for {3D} classification and
  segmentation.
\newblock In {\em Proceedings of the IEEE conference on computer vision and
  pattern recognition}, pages 652--660, 2017.

\bibitem{liu2022fluid}
Q.~Liu, W.~Zhu, X.~Jia, F.~Ma, and Y.~Gao.
\newblock Fluid simulation system based on graph neural network.
\newblock {\em arXiv:2202.12619}, 2022.

\bibitem{gruber2022comparison}
A.~Gruber, M.~Gunzburger, L.~Ju, and Z.~Wang.
\newblock A comparison of neural network architectures for data-driven
  reduced-order modeling.
\newblock {\em Comput. Methods Appl. Mech. Eng.}, 393:114764, 2022.

\bibitem{GSW2020}
H.~Gao, L.~Sun, and J.-X. Wang.
\newblock {P}hy{G}eo{N}et: Physics-informed geometry-adaptive convolutional
  neural networks for solving parameterized steady-state pdes on irregular
  domain.
\newblock {\em J. Comput. Phys.}, page 110079, 2020.

\bibitem{Kajishima}
T.~Kajishima and K.~Taira.
\newblock {\em Computational {F}luid {D}ynamics: {I}ncompressible {T}urbulent
  {F}lows}.
\newblock Springer, 2017.

\bibitem{kochkov2021machine}
D.~Kochkov, J.~A. Smith, A.~Alieva, Q.~Wang, M.~P. Brenner, and S.~Hoyer.
\newblock Machine learning--accelerated computational fluid dynamics.
\newblock {\em Proc. Natl. Acad. Sci. U.S.A.}, 118(21):e2101784118, 2021.

\bibitem{stolz2001approximate}
S.~Stolz, N.~A. Adams, and L.~Kleiser.
\newblock An approximate deconvolution model for large-eddy simulation with
  application to incompressible wall-bounded flows.
\newblock {\em Phys. Fluids}, 13(4):997--1015, 2001.

\bibitem{du2023state}
Y.~Du, M.~Wang, and T.~A. Zaki.
\newblock State estimation in minimal turbulent channel flow: {A} comparative
  study of {4DVar} and {PINN}.
\newblock {\em Int. J. Heat Fluid Flow}, 99:109073, 2023.

\bibitem{di2020synchronization}
P.~C. Di~Leoni, A.~Mazzino, and L.~Biferale.
\newblock Synchronization to big data: {N}udging the {N}avier-{S}tokes
  equations for data assimilation of turbulent flows.
\newblock {\em Phys. Rev. X}, 10(1):011023, 2020.

\bibitem{di2018inferring}
P.~C. Di~Leoni, A.~Mazzino, and L.~Biferale.
\newblock Inferring flow parameters and turbulent configuration with
  physics-informed data assimilation and spectral nudging.
\newblock {\em Phys. Rev. Fluids}, 3(10):104604, 2018.

\bibitem{yasuda2022spatio}
Y.~Yasuda and R.~Onishi.
\newblock Spatio-temporal super-resolution data assimilation ({SRDA}) utilizing
  deep neural networks with domain generalization technique toward
  four-dimensional {SRDA}.
\newblock {\em arXiv:2212.03656}, 2022.

\bibitem{yousif2022physics}
M.~Z. Yousif, L.~Yu, and H.-C. Lim.
\newblock Physics-guided deep learning for generating turbulent inflow
  conditions.
\newblock {\em J. Fluid Mech.}, 936:A21, 2022.

\bibitem{nakamura2022identifying}
T.~Nakamura, K.~Fukami, and K.~Fukagata.
\newblock Identifying key differences between linear stochastic estimation and
  neural networks for fluid flow regressions.
\newblock {\em Sci. Rep.}, 12(3726), 2022.

\bibitem{Fukamipump2022}
K.~Fukami, B.~An, M.~Nohmi, M.~Obuchi, and K.~Taira.
\newblock Machine-learning-based reconstruction of turbulent vortices from
  sparse pressure sensors in a pump sump.
\newblock {\em J. Fluids Eng.}, 144(12):121501, 2022.

\bibitem{li2008public}
Y.~Li, E.~Perlman, M.~Wan, Y.~Yang, C.~Meneveau, R.~Burns, S.~Chen, A.~Szalay,
  and G.~Eyink.
\newblock A public turbulence database cluster and applications to study
  {L}agrangian evolution of velocity increments in turbulence.
\newblock {\em J. Turb}, (9):N31, 2008.

\bibitem{wu2008direct}
X.~Wu and P.~Moin.
\newblock A direct numerical simulation study on the mean velocity
  characteristics in turbulent pipe flow.
\newblock {\em J. Fluid Mech.}, 608:81--112, 2008.

\bibitem{towne2022database}
A.~Towne, S.~Dawson, G.~A. Br{\`e}s, A.~Lozano-Dur{\'a}n, T.~Saxton-Fox,
  A.~Parthasarathy, A.~R. Jones, H.~Biler, C.-A. Yeh, H.~D. Patel, and
  K.~Taira.
\newblock A database for reduced-complexity modeling of fluid flows.
\newblock {\em AIAA J.}, 2023.

\end{thebibliography}
%%% Remove comment to use the external .bib file (using bibtex).

\end{document}